\documentclass{article} 
\usepackage{iclr2024_conference,times}

\usepackage{amsmath,amsfonts,bm}

\def\eqref#1{equation~\ref{#1}}

\def\1{\bm{1}}

\def\rve{{\mathbf{e}}}

\def\rvg{{\mathbf{g}}}
\def\rvh{{\mathbf{h}}}

\def\rvk{{\mathbf{k}}}

\def\rvo{{\mathbf{o}}}

\def\rvq{{\mathbf{q}}}

\def\rvs{{\mathbf{s}}}

\def\rvv{{\mathbf{v}}}
\def\rvw{{\mathbf{w}}}

\def\rmH{{\mathbf{H}}}

\def\rmK{{\mathbf{K}}}

\def\rmQ{{\mathbf{Q}}}

\def\rmT{{\mathbf{T}}}

\def\rmW{{\mathbf{W}}}

\def\vc{{\bm{c}}}

\def\ve{{\bm{e}}}

\def\vp{{\bm{p}}}

\def\vu{{\bm{u}}}

\DeclareMathAlphabet{\mathsfit}{\encodingdefault}{\sfdefault}{m}{sl}
\SetMathAlphabet{\mathsfit}{bold}{\encodingdefault}{\sfdefault}{bx}{n}

\def\gE{{\mathcal{E}}}

\def\gG{{\mathcal{G}}}

\def\gQ{{\mathcal{Q}}}

\def\gS{{\mathcal{S}}}
\def\gT{{\mathcal{T}}}

\newcommand{\R}{\mathbb{R}}

\usepackage{hyperref}
\usepackage{url}

\usepackage{booktabs}       
\usepackage{amsfonts}       
\usepackage{nicefrac}       
\usepackage{microtype}      
\usepackage{xcolor}         

\usepackage{subfloat}
\usepackage{float}
\usepackage{graphicx}
\usepackage{comment}
\usepackage{amsmath,amssymb} 
\usepackage{color}
\usepackage{colortbl}
\usepackage{bbding}
\usepackage{xspace}
\usepackage{textcomp}
\usepackage{mathtools}
\usepackage{subfig}
\usepackage{multirow}
\usepackage{pifont}
\usepackage{mathrsfs}
\usepackage{algorithm}
\usepackage{algorithmic}
\usepackage{multirow}
\usepackage{amssymb}
\usepackage{booktabs} 
\usepackage{graphicx}
\usepackage{tabulary}
\usepackage{wrapfig}
\usepackage{diagbox}
\usepackage{setspace}

\definecolor{lm_purple}{RGB}{227,227,240}

\usepackage[normalem]{ulem}
\useunder{\uline}{\ul}{}

\title{De novo Protein Design Using Geometric \\ Vector Field Networks}

\author{%
Weian Mao$ ^{1,2}  \thanks{WM, ZS and MZ contributed equally. Work was done when WM was visiting Zhejiang University.
}$, \quad
Muzhi Zhu$^{1} \footnotemark[1]$,\quad
Zheng Sun$^{3} \footnotemark[1]$,\quad
Shuaike Shen$^1$,\\
\bf Lin Yuanbo Wu$^3$,\quad
Hao Chen$^1$,\quad
Chunhua Shen$^1$\\[.122cm]
  $^1$ Zhejiang University, China
  ~~~
  ~~~
  $^2$ The University of Adelaide, Australia
  \\
  $^3$ Swansea University, UK
}

\iclrfinalcopy 
\begin{document}

\maketitle

\begin{abstract}

Innovations like protein diffusion have enabled significant progress in \textit{de novo} protein design, which is a vital topic in life science.
These methods typically depend on protein structure encoders to model residue backbone frames, where atoms do not exist. Most prior encoders rely on atom-wise features, such as angles and distances between atoms, which are not available in this context. Thus far, only 
several simple 
encoders, 
such as 
IPA \citep{jumper2021highly}, have been proposed for this scenario, exposing the frame modeling as a bottleneck. In this work, we 
proffer 
the Vector Field Network (VFN), 
which 
enables network layers to perform learnable vector computations between coordinates of frame-anchored virtual atoms, thus achieving a higher capability for modeling frames. The vector computation operates in a manner similar to a linear layer, with each input channel receiving 3D virtual atom coordinates instead of scalar values. The multiple feature vectors output by the vector computation are then used to update the residue representations and virtual atom coordinates via attention aggregation. Remarkably, VFN also excels in modeling both frames and atoms, as the real atoms can be treated as the virtual atoms for modeling, positioning VFN as a potential \textit{universal encoder}. In protein diffusion (frame modeling), VFN exhibits an impressive performance advantage over IPA, excelling in terms of both designability (\textbf{67.04}\% vs.\ 53.58\%) and diversity (\textbf{66.54}\% vs. 51.98\%). In inverse folding (frame and atom modeling), VFN outperforms the previous SoTA model, PiFold (\textbf{54.7}\% vs.\ 51.66\%), on sequence recovery rate. We also propose a method of equipping VFN with the ESM model 
\citep{lin2022language}, which significantly surpasses the previous ESM-based SoTA (\textbf{62.67}\% vs.\  55.65\%), LM-Design \citep{zheng2023lm_design}, by a substantial margin.
\end{abstract}
%

\begin{figure}[h]
\vspace{-2.0em}
    \centering
    \includegraphics[width=1.0\linewidth]{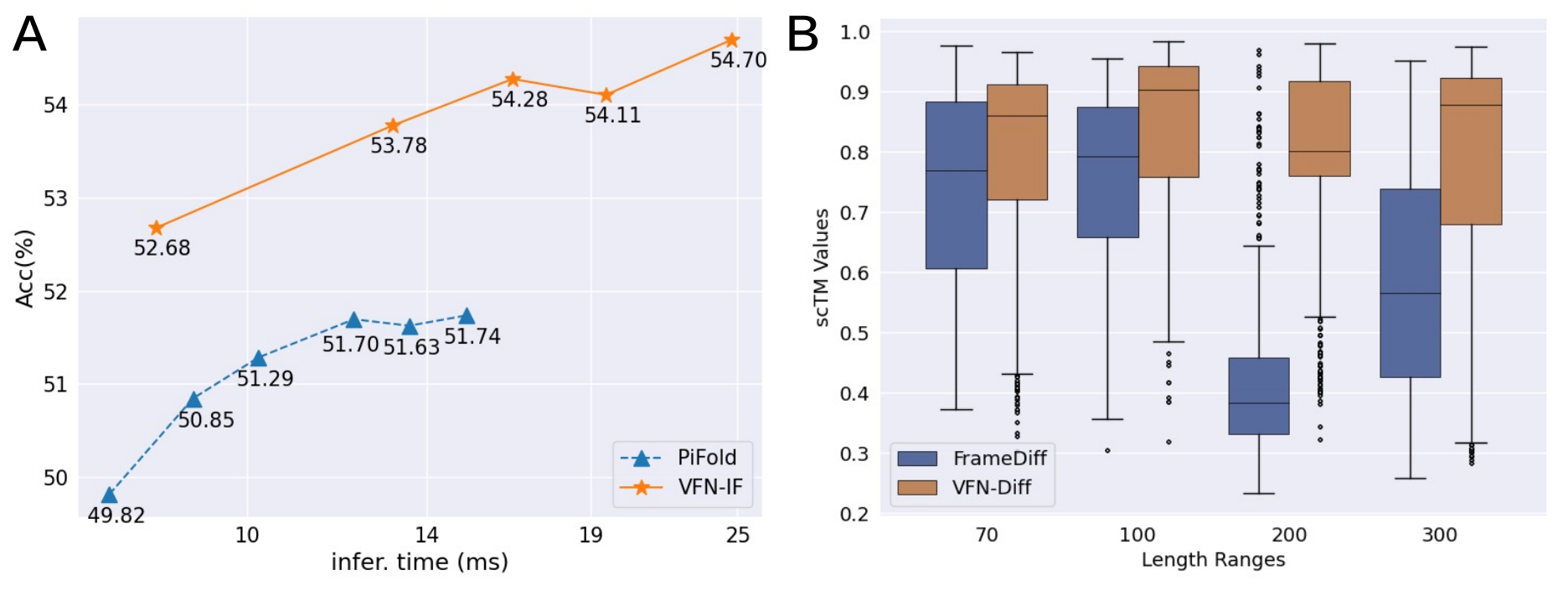}
    \vspace{-2.5em}
    \caption{\small{Experimental results on protein diffusion and inverse folding. A) VFN-IF compared to PiFold with varying numbers of layers, showcasing the trade-off between speed and sequence recovery rate. B) VFN-Diff compared to FrameDiff (IPA) for designability across proteins of varying lengths.}}\label{fig:first_page}
    \label{fig:fig_at_first_page}
    \vspace{-1.0em}
\end{figure}

\section{Introduction}
The field of \textit{de novo} protein design        \citep{huang2016coming} represents a pivotal frontier in the realms of bioengineering and drug development, holding the potential to bring about a revolutionary shift in the creation of innovative therapeutic agents. Recent advancements in this domain have been driven by a groundbreaking paradigm that combines protein structure diffusion models        \citep{watson2023novo, yim2023se} with inverse folding networks        \citep{dauparas2022robust}. Specifically, this paradigm initiates by employing a protein diffusion model to stochastically generate the backbone structure of a protein, represented by residue frames. Since the types of amino acids in the generated protein are initially unknown, an inverse folding network is then utilized to design the protein sequences for each residue based on the backbone residue frames. While this paradigm brings huge success for protein design, it also brings challenges for deep learning-based protein structure encoders.

Methods like protein diffusion rely on protein structure encoders to model and sample protein structures. However, prior structure encoders are unsuitable or severely limited in this case, primarily because the representation of residues is highly specialized in methods like diffusion. Specifically, in those methods, backbone frames are commonly employed to represent the spatial information of residues, with atom-level representations typically absent. The absence of atoms has rendered most previous encoders ineffective for protein design, as they typically rely on atom-level input features such as interatomic angles and distances. Although a few basic encoders, such as IPA        \citep{jumper2021highly}, were designed for frame modeling, they still faced significant limitations. For instance, IPA simply performs distance pooling between frame-anchored virtual atoms as geometric features. Obviously, such pooling operation and the single pooled distance value clearly lacks the expressiveness required for representation. We refer to these limitations as the \textit{atom representation bottleneck}, which is explained in detail in Section~\ref{sec:design_encoder}. In response to this challenge, we propose a novel structure encoder called the Vector Field Network (VFN).

The core idea of VFN revolves around the utilization of a vector-specific linear layer to extract multiple geometric feature vectors by mapping the coordinates of frame-anchored virtual atoms. Specifically, VFN introduces virtual atoms in Euclidean space for each amino acid, and these virtual atoms move in conjunction with the frames, serving as dynamic representations of the frames. When modeling the relationship between two amino acids, a module called the `vector field operator' takes the virtual atom coordinates of the two amino acids as vector inputs and performs operations similar to a linear layer. Within the vector field operator, each vector is initially multiplied by a learnable scalar weight, and these weighted vectors are accumulated to compute the output feature vector. Like a linear layer, the vector field operator has multiple output channels, with each output channel generating a Euclidean vector. Subsequently, these output feature vectors are directed into an MLP layer for processing, facilitating the fusion of residue features and enabling the modeling of residue frames.

VFN exhibits significant advantages over previous encoders in terms of its expressive power and versatility, as VFN circumvents the \textit{atom representation bottleneck} present in IPA. As previously mentioned, IPA relies on a single scalar variable, which represents the sum of distances between virtual atoms, serving as both geometric features and attention bias. This single scalar variable and corresponding pooling operation limit the expressive capacity of IPA. In contrast, VFN can flexibly extract multiple feature vectors through a vector field operator, thereby circumventing this bottleneck. Please refer to section~\ref{sec:design_encoder} for more details.

In scenarios where frame and backbone atoms coexist, such as inverse folding, VFN maintains excellent expressiveness while simultaneously offering enhanced generality and flexibility. This is because real atoms can be treated as virtual atoms in VFN, and the coordinates of both real and virtual atoms can be used to model the relationships between amino acids using the vector field operator, naturally forming hierarchical representations. 
Compared to other atom-based inverse folding methods, VFN achieves superior performance by potentially modeling frames more effectively. Thus, VFN also surpasses the current SoTA in inverse folding. It is worth emphasizing that, IPA is not suitable for treating the real atoms as the virtual atoms, since the IPA network cannot provide learnable weights for the coordinates of each atom, facing the \textit{atom representation bottleneck}.

To assess the performance of VFN for \textit{de novo} protein design, we have implemented two models based on VFN, namely VFN-IF and VFN-Diff, tailored for protein inverse folding and diffusion based generative modeling, respectively. Experimental results consistently demonstrate the remarkable performance of VFN. For protein diffusion, VFN-Diff significantly outperforms the prior solid baseline, FrameDiff        \citep{yim2023se}, in terms of designability (\textbf{67.04}\% \textit{vs.} 53.58\%) and diversity (\textbf{66.54}\% \textit{vs.}\  51.98\%). It is important to emphasize that the key distinction between VFN-Diff and FrameDiff lies in the replacement of the encoder—IPA, with VFN layers. This substantiates the superior geometric feature extraction capability of VFN over the widely adopted IPA.

In the domain of inverse folding, VFN-IF exhibits a substantial performance boost over the current state-of-the-art method, PiFold (\textbf{54.74}\% \textit{vs.} 51.66\%, sequence recovery rate). Furthermore, we have trained a larger-scale VFN-IF model on the entire PDB database, achieving an impressive sequence recovery rate of 57.14\%, underscoring the scalability of VFN-IF. 
Additionally, we propose a novel variant of VFN-IF, called VFN-IFE, which is equipped with an external knowledge base, achieves remarkable precision at \textbf{62.64}\%, surpassing SoTA approaches in this regard, LM-Design \citep{zheng2023lm_design} (55.65\%), by a substantial margin.

The main contributions of this work can be summarized as follows:
\begin{itemize}
\item We 
propose 
the Vector Field Network (VFN), which employs a vector field operator to extract geometric feature vectors between frame-anchored virtual atoms, resembling a linear layer. This approach overcomes the \textit{atom representation bottleneck}, thereby enhancing representational capabilities. Notably, VFN can also incorporate real atoms as the virtual atoms for hierarchical modeling, positioning it as a potential \textit{universal encoder}.

\item For protein diffusion, VFN significantly enhances the designability(\textbf{67.04}\% \textit{vs.} 53.58\%) and diversity(\textbf{66.54}\% \textit{vs.} 51.98\%) of protein generation compared to IPA.

\item For inverse folding, VFN significantly surpasses the previous SoTA model, PiFold (\textbf{54.74}\% \textit{vs.} 51.66\%, sequence recovery rate). Additionally, we propose a method for equipping VFN with an external knowledge base, achieving a substantial breakthrough over the SoTA approach, LM-Design, in this regard, thus elevating model accuracy to next level (\textbf{62.64}\% \textit{vs.} 55.65\%).

\end{itemize}

\section{Related Work}
\subsection{De novo Protein Design}
\vspace{-0.5em}
\textit{De novo} protein design, which involves the creation of proteins from scratch, holds paramount significance in the fields of enzyme engineering and protein engineering. Traditional approaches such as RosettaDesign        \citep{liu2006rosettadesign} were prevalent before the advent of machine learning-based methods. In recent years, with the maturation of machine learning techniques, advanced deep learning-based methods have emerged, exemplified by \citet{huang2022backbone} (Side Chain Unknown Backbone Arrangement) and protein diffusion        \citep{watson2023novo,fu2023latent,gao2023diffsds}.

The paradigm of protein diffusion is regarded as one of the most promising methods in the realm of protein design, which encompasses protein diffusion        \citep{yim2023se} and inverse folding        \citep{gao2022pifold, jendrusch2021alphadesign, wu2021protein, ovchinnikov2021structure, dauparas2022robust, ingraham2019generative, hsu2022learning, gao2023knowledge, derevyanko2018deep}. Specifically, a protein diffusion model first generates the backbone structure of a protein, followed by an inverse folding network that designs the corresponding sequence for this backbone. The feasibility of both these steps has been experimentally validated through cryo-electron microscopy        \citep{watson2023novo, dauparas2022robust}, marking a significant breakthrough in the field of protein design. 
However, while protein diffusion methods based on frame representation achieve significant success, in these methods, atom representation is absent, rendering previous general purpose encoders unusable.

\vspace{-0.5em}
\subsection{General Purpose Encoder}
\vspace{-0.5em}
In the past, numerous encoders        \citep{hermosilla2020intrinsic, zhang2022protein, hermosilla2022contrastive, velivckovic2017graph, baldassarre2021graphqa,li2022directed,gao2022alphadesign, shroff2019structure, dumortier2022petribert, mcpartlon2022deep, cao2021fold2seq, anishchenko2021novo, karimi2020novo, zhang2020prodconn, wang2022comenet, derevyanko2018deep} have been proposed for tasks such as model quality assessment        \citep{townshend2021atom3d} and fold classification        \citep{hou2018deepsf}, where atomic information is available. However, these methods are not suitable for protein design tasks where atomic representations of proteins are unavailable. 
For instance, GVP        \citep{jing2020learning} transforms input atomic coordinates into vector and scalars variables as the network input, facilitating the model's SE(3) invariance. 
Meanwhile,         \citet{wang2022learning, jin2022antibody} proposed efficient approaches for modeling protein structures using hierarchical protein representations, enabling GNNs to possess both residue-level and atom-level hierarchical representations. Nevertheless, such approaches are evidently less applicable in the methods like protein diffusion, as atomic information is unattainable.
\vspace{-0.5em}
\subsection{Frame-based Encoder}
\vspace{-0.5em}
The use of frame-based representations in protein structure encoders has been relatively unexplored, with few existing encoders being rather rudimentary in nature. Historically, these encoders have primarily served as auxiliary modules in protein structure prediction models, such as RoseTTAFold        \citep{baek2021accurate} and AlphaFold2        \citep{jumper2021highly}. In RoseTTAFold, the explicit concept of frames is absent; instead, it employs the origin of frames to represent frames and employs SE(3)-transformers        \citep{fuchs2020se} to model these frames. However, this approach conspicuously neglects rotational information of frames, considering only positional offsets. AlphaFold2, on the other hand, introduces the IPA structural encoding module to address this limitation. IPA represents frames using framed-anchored virtual atoms and extracts geometric information between two frames by computing the sum of distances between virtual atom pairs. This approach takes into account both the coordinate offsets between frames and their rotational information.
\vspace{-0.5em}
\subsection{Atom Representation Bottleneck}\label{sec:design_encoder}
\vspace{-0.5em}
However, IPA still faces the \textit{atom representation bottleneck}. Firstly, IPA does not incorporate learnable weights when extracting features for virtual atoms; it directly applies a simple summation pooling operation to the atom-pair distances, thereby lacking flexibility. Secondly, the sum of distances between virtual atom pairs yields only a scalar variable, making it challenging to represent complex geometric information, thus lacking expressiveness. In VFN, these issues have been successfully resolved. VFN introduces an operator similar to a linear layer, characterized by the inclusion of learnable weights associated with individual atom coordinates. These weights facilitate flexible vector computations, effectively mitigating the flexibility challenges inherent in IPA's pooling operation. Furthermore, the operator in VFN can return multiple feature vectors, eliminating the expressiveness bottleneck caused by the single scalar variable (the pooled distance) in the IPA. For further details regarding the vector field operator, please refer to section~\ref{sec: vector_field_operator}.

\section{Vector Field Network Layers}
\vspace{-0.5em}


The Vector Field Network (VFN) is designed to extract geometric features between amino acids using a module, named the vector field operator~\ref{sec: vector_field_operator}. In each layer of VFN, the protein's representation undergoes a sequential process including the vector field operator, node interactions~\ref{sec:node_inter}, edge interactions~\ref{sec:edge_inter} and virtual atom updating~\ref{sec:edge_inter}. The vector field operator is crucial for extracting geometric features between pairs of amino acids through vector computations and virtual atoms. Subsequently, these extracted geometric features are aggregated and employed to update the representation of each amino acid through node and edge interactions. At the end of each layer, the coordinates of the mentioned virtual atom are updated by aggregation or residue representations. The mentioned modules and the overall pipeline are elucidated subsequently.
\vspace{-0.5em}
\subsection{Protein Representation}
\vspace{-0.5em}
In VFN, a protein consisting of $n$ amino acids is represented as a graph denoted as $\gG = (\gS, \gE, \gT,\gQ)$. Here, $\gS = \{\rvs_i \in \R^{d_v}\}_{i=1,...,n}$ represents the set of node features. To encode the positional information of each amino acid in space, we use a set of local frames, $\gT= \{\rmT_i\}_{i=1,...,n}$, to represent the position of each amino acid. We introduce a set of frame-anchored virtual atoms, whose coordinates are maintained by Eq.~\ref{eq:s_to_atom} or Eq.~\ref{eq:attn_to_atom}. Specifically, we denote all virtual atom coordinates as $\gQ= \{\rmQ_i\}_{i=1,...,n}$, with $\rmQ_i=\{\vec{\rvq}_{k}\in \R^3\}_{k=1,...,d_q}$ representing a set of virtual atom coordinates associated with the $i$-th residue \textit{w.r.t.} $\rmT_i$. $d_q$ denotes the number of virtual atoms in each residue. These virtual atom coordinates are treated as vectors in the vector field operator, allowing for vector computations to extract geometric features. Additionally, we construct edges in the graph based on specific rules, such as a complete graph or $k$-nearest neighbors. The complete graph is taken by default, so the set of edge features can be denoted as $\gE = \{\rve_{i,j}\in \R^{d_e}\}_{i,j=1,...,n}$.

\begin{figure}[h]
\vspace{-1.0em}
    \centering
    \includegraphics[width=0.8\linewidth]{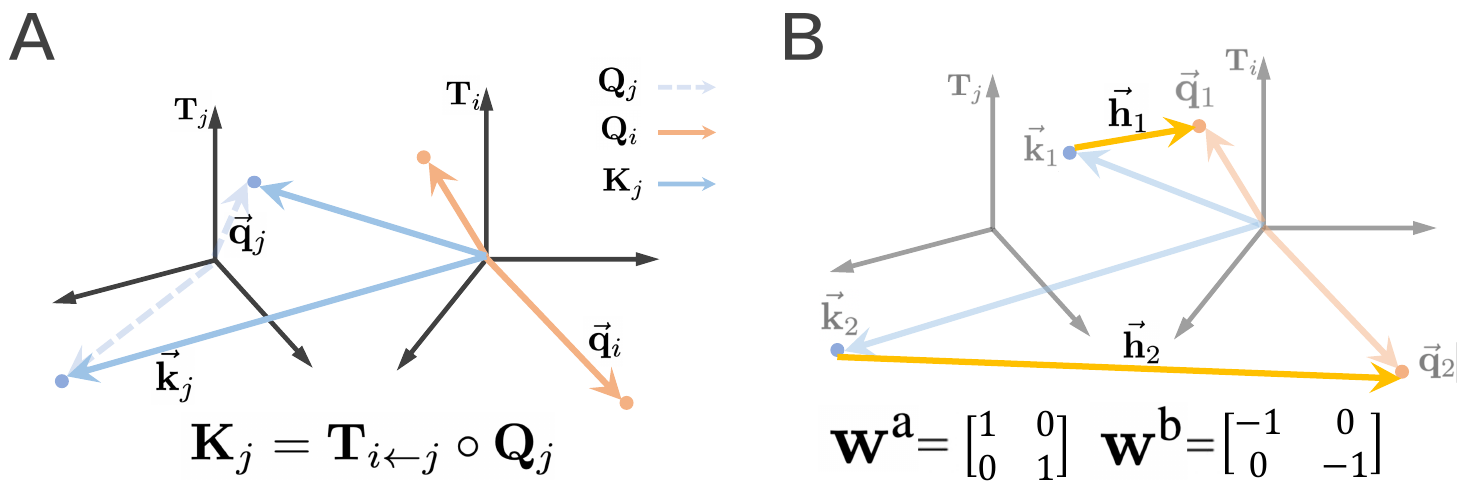}
    \vspace{-1.0em}
    \caption{Pipeline for the Vector Field Operator. A) Transforming the virtual atomic coordinates $\rmQ_j$ from frame $\rmT_j$ to frame $\rmT_i$ to obtain $\rmK_j$. B) An example of vector computation involving vectors $\rmQ_i$ and $\rmK_j$ using learnable weights $\rvw^\text{a}$ and $\rvw^\text{b}$ as defined in Equation~\ref{eq:vector_culculation}. When $\rvw^\text{a}$ and $\rvw^\text{b}$ are specific weights (as shown in figure), the vector field can yield the Euclidean vector, $\vec{\rvh}_{1}$ and $\vec{\rvh}_{2}$, between two particular atoms.}
    \label{fig:vector_field_oprator}
    \vspace{-1.0em}
\end{figure}
\vspace{-0.5em}
\subsection{Vector Field Operator}\label{sec: vector_field_operator}
\vspace{-0.5em}
As shown in figure~\ref{fig:vector_field_oprator}, the vector field operator extracts geometric features between two amino acids by performing learnable vector calculations on their virtual atom coordinates. Specifically, when modeling the geometric relationship between two amino acids, such as the $i$-th and $j$-th amino acids, the input to the vector field operator consists of virtual atom groups, $\rmQ_i$ and $\rmQ_j$. To ensure that all vectors are represented in the same local coordinate system, we first transform $\rmQ_j$ from the local frame $\rmT_j$ to $\rmT_i$, and denote the transformed coordinates as $\rmK_j=\{\vec{\rvk}_{k} \in \R^3\}_{k=1,...,d_q}$, which can be written as (as shown in figure~\ref{fig:vector_field_oprator}.A ):
\begin{equation}\label{eq:transform_op}
    \rmK_j = \rmT_{i\leftarrow j} \circ \rmQ_j, \quad  \text{where} \ \  \rmT_{i\leftarrow j} = \rmT_i^{-1} \circ \rmT_j
\end{equation}
Here, $\rmT_{i\leftarrow j}$ represents the transformation matrix transforming the coordinates from $\rmT_i$ to $\rmT_j$. Next, to select specific virtual atoms for calculating feature vectors, similar to a linear layer, we introduce two sets of learnable weights, $\rvw^\text{a}=\{w^\text{a}_{k,l} \in \R \}_{k,l=1,...,d_q}$ and $\rvw^\text{b}=\{w^\text{b}_{k,l} \in \R \}_{k,l=1,...,d_q}$. Those weights are then utilized to perform vector calculations between $\rmQ_i$ and $\rmK_j$, resulting in extracted feature vectors $\rmH_{i,j}=\{\vec{\rvh}_{k} \in \R^3\}_{k=1,...,d_q}$ (as shown in figure~\ref{fig:vector_field_oprator}.B ). This can be expressed as follows:
\begin{equation}\label{eq:vector_culculation}
\vec{\rvh}_{k} = \sum_l w^\text{a}_{k,l} \vec{\rvq}_{l} + \sum_l w^\text{b}_{k,l} \vec{\rvk}_{l}
\end{equation}

where $\vec{\rvq}_{l} \in \rmQ_i, \vec{\rvk}_{l} \in \rmK_j, \vec{\rvh}_{k} \in \rmH_{i,j}$. Here, $\rmH_{i,j}$ serves as a vector representation for captured geometric features. However, due to its large numerical range (ranging from $-200$\AA\  to $200$\AA), it can lead to instability during network training and requires further processing. 

To avoid this issue, $\rmH_{i,j}$ is decomposed into two variables: unit direction vectors and vector lengths. The vector lengths will be mapped using a radial basis function, $\operatorname{RBF}$. This can be written as:
\begin{equation}\label{eq:h_to_g}
\rvg_{i,j} = \operatorname{concat}_k( \frac{\vec{\rvh}_{k}}{\Vert \vec{\rvh}_{k} \Vert}, \operatorname{RBF}(\Vert \vec{\rvh}_{k} \Vert)),  \quad \rvg_{i,j} \in \R^{d_g} 
\end{equation}
$\rvg_{i,j}$ is a vector that represents the geometric relationship between two residues and is used in the following module for aggregating and updating the features $\rvs_i$ and $\rve_{i,j}$. The $\operatorname{concat}_k$ represents the concatenation of all the feature\footnote{representing the flattened unit direction vectors and the values output by $\operatorname{RBF}$} resulting from all the vectors in $\rmH_{i,j}=\{\vec{\rvh}_{k}\}_{k=1,...,d_q}$. $d_g$ represents the number of channels in $\rvg_{i,j}$.
\vspace{-0.5em}
\subsection{Node Interactions}\label{sec:node_inter}
\vspace{-0.5em}
Here, a MLP-based multi-head attention mechanism is designed to aggregate geometric features $\rvg_{i,j}$, node features $\rvs_i$ and $\rvs_j$, edge features $\rve_{i,j}$, and update the node representations $\rvs_i$. Specifically, the pair-wise features mentioned above are first fed into an MLP, followed by a softmax operation to obtain attention weights, denoted as:
\begin{equation}\label{eq:attn}
    a_{i,j} = \operatorname{softmax}_j(\operatorname{MLP}(\rvs_i, \rvs_j, \rvg_{i,j}, \rve_{i,j}))
\end{equation}
Where $a_{i,j}$ represents the attention weight for the interaction between nodes $i$ and $j$. Next, another MLP is employed to generate the values $\rvv_{i,j}$ for the multi-head attention, which are subsequently aggregated using the attention mechanism, expressed as follows:
\begin{equation}
    \rvo_{i} = \sum_j a_{i,j} \rvv_{i,j}, \quad \text{where} \ \  \rvv_{i,j} = \operatorname{MLP}(\rvs_j, \rvg_{i,j}, \rve_{i,j})
\end{equation}
Here, $\rvo_{i}$ represents the aggregated features, which are utilized to update the features $\rvs_i$ after undergoing an MLP layer, written as:
\begin{equation}
    \rvs_i \leftarrow \rvs_i + \operatorname{MLP}(\rvo_{i})
\end{equation}
Up to this point, node features $\rvs_i$ have been updated and are utilized as the input for the subsequent layer and following operations.
\vspace{-0.5em}
\subsection{Edge Interactions}\label{sec:edge_inter}
\vspace{-0.5em}
Next, we introduce the edge interactions, which is designed to aggregate geometric information $\rvg_{i,j}$, node features $\rvs_i$ and $\rvs_j$, and edge features $\rve_{i,j}$, to update the representation of the edge $\rve_{i,j}$. This can be written as follows:
\begin{equation}\label{eq:edge_inter}
    \rve_{i,j} \leftarrow \rve_{i,j} + \operatorname{MLP}(\rvs_i, \rvs_j, \rvg_{i,j}, \rve_{i,j})
\end{equation}
\vspace{-0.5em}
\subsection{Virtual Atom Coordinates Updating}\label{sec:atom_update}
\vspace{-0.5em}
In the final stage of each VFN layer, the coordinates of virtual atoms $\rmQ_i$ are updated. We have devised two different methods for updating these virtual atom coordinates, which are respectively referred to as `node feature-based updating' and `coordinate aggregating updating.' These two approaches can be selected based on the specific task, and their detailed methodologies are elucidated following.

\textbf{Node Feature-Based Updating}. Node features $\rvs_i$ are processed through a linear layer to generate a set of virtual atom coordinates for updating $\rmQ_i$. This process can be represented as follows:
\begin{equation}\label{eq:s_to_atom}
    \rmQ_i \leftarrow \operatorname{Linear}(\rvs_i)
\end{equation}

\textbf{Coordinate Aggregating Updating}. For updating $\rmQ_i$, virtual atom coordinates $\rmK_j$ are firstly aggregated through an attention mechanism to obtain the aggregated atom coordinates $\rmQ_i^\text{o}$. Subsequently, $\rmQ_i^o$ is fed into an MLP layer to update the coordinates $\rmQ_i$, denoted as follows:
\begin{equation}\label{eq:attn_to_atom}
\begin{aligned}
    \rmQ_i \leftarrow \operatorname{V-MLP}(\rmQ_i, \rmQ_i^\text{o}), \quad \text{where} \ \ \rmQ_i^\text{o} = \sum_j a_{i,j} \rmK_j
\end{aligned}
\end{equation}
Here, $\operatorname{V-MLP}$ represents a dedicated MLP layer designed specifically for vectors, as described in the appendix~\ref{sec:vmlp}; $a_{i,j}$ is computed in equation \ref{eq:attn}.

\begin{wrapfigure}{r}{0.55\columnwidth}
    \vspace{-1.0em}
    \includegraphics[width=0.55\columnwidth]{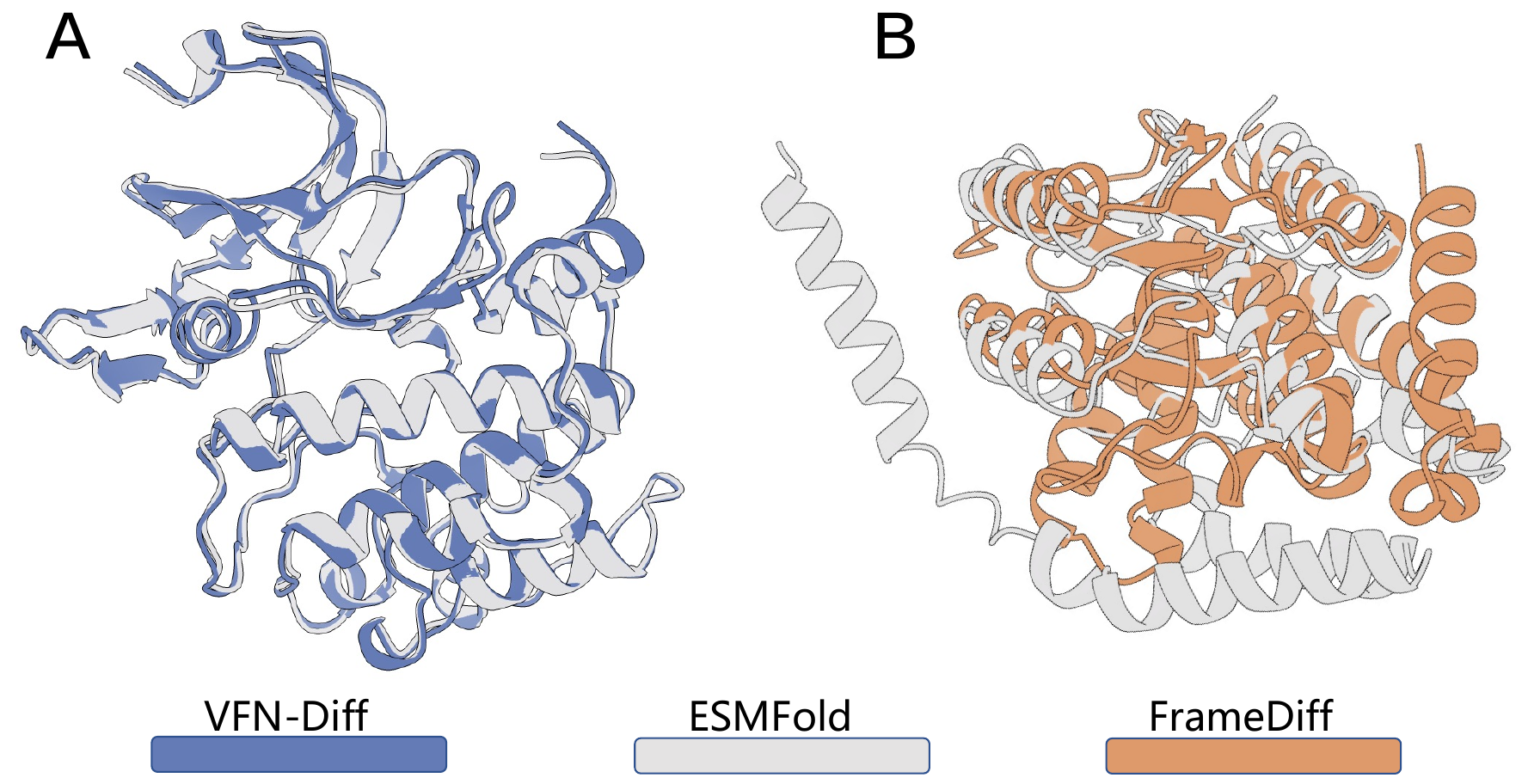}
    \vspace{-1.0em}
    \caption{\small Visual Comparison of VFN-IF and Frame-Diff.  `ESMFold' represents protein structures recovered using ProteinMPNN and ESMFold, with closer structural resemblance being preferable.}\label{fig:diffusion_vis}
    \vspace{-1.0em}
\end{wrapfigure}

\section{Implementation for De novo Protein Design}
To establish the recent paradigm in \textit{de novo} protein design, we have developed two distinct models, namely VFN-Diff and VFN-IF, each dedicated to protein structure diffusion and inverse folding, respectively. In the protein diffusion part, the protein structure is designed and represented using backbone frames $\gT$. Subsequently, these backbone frames are fed into the inverse folding network to obtain the corresponding protein sequence for the designed structure. In the following subsections, we present an overview of the implementations for VFN-Diff and VFN-IF. For more implementation details, please refer to the appendix~\ref{sec:vfn_diff} and appendix~\ref{sec:vfn_if}.
\vspace{-0.5em}
\subsection{Protein Diffusion}
\vspace{-0.5em}

We adopted the FrameDiff        \citep{yim2023se} paradigm, a diffusion model used to sample protein backbone structures by updating residue frames $\gT$. In FrameDiff, a network called $\operatorname{FramePred}$ is employed to model the protein backbone frames $\gT$ during each diffusion step. This network relies on invariant point attention (IPA), which consists of three components: node attention, edge attention, and point attention. Importantly, the point attention module is the operation that causes the \textit{atom representation bottleneck}, as mentioned in section~\ref{sec:design_encoder}. Therefore, to fairly evaluate whether VFN can bypass this bottleneck, we replaced the point attention module with our VFN layer. The remaining parts of VFN-Diff remain consistent with FrameDiff.

\subsection{Inverse Folding}
\begin{wrapfigure}{r}{0.55\columnwidth}
    \vspace{-2.5em}
    \includegraphics[width=0.55\columnwidth]{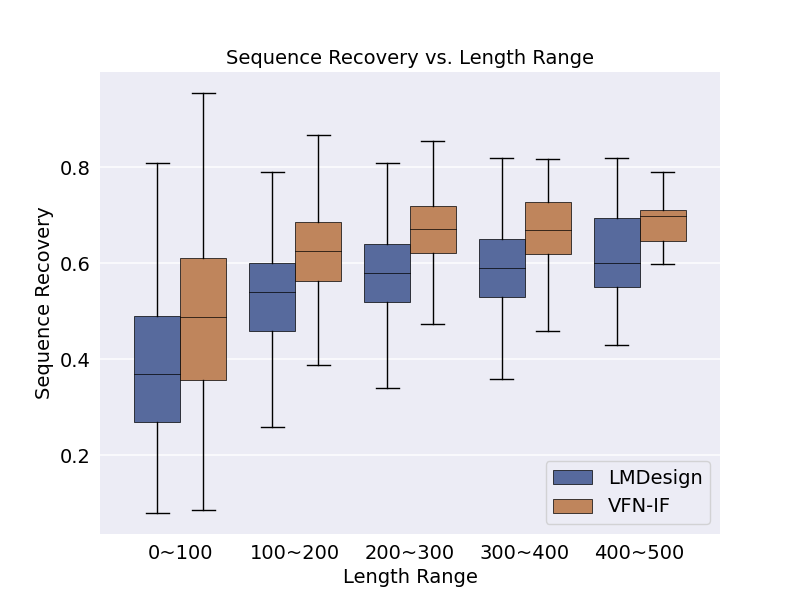}
    \vspace{-2.0em}
    \caption{Visualization comparison, conducted on inverse folding and CATH 4.2, between sequence recovery rate and protein length.}\label{fig:if_acc_length}
    \vspace{-1.0em}
\end{wrapfigure}
The purpose of the inverse folding task is to map the frames $\gT$ (generated by the diffusion model) to amino acid categories $\vc \in \{1,...,20\}^n$, denoted as $\displaystyle f_{\text{if}}: \gT \rightarrow \vc$. These predicted amino acid categories aim to enable the protein to fold into the designed structure. 
In this task, we adopt the previous paradigm. Specifically, the VFN-IF network is composed of 15 sequential VFN layers, and it constructs edges in the graph using a $k$-NN approach. The network takes backbone frames and backbone atoms as input. It's important to note that the coordinates of backbone atoms are used to initialize some of the virtual atom coordinates, enabling VFN to achieve higher modeling capacity through hierarchical representation. 
The prediction head of VFN-IF follows the PiFold approach        \citep{gao2022pifold}, directly predicting amino acid categories for each node through a linear layer and supervising them using a cross-entropy loss function.
Additionally, we propose an approach to fine-tune the ESM model        \citep{lin2022language} with LoRA        \citep{hu2021lora} to correct the predictions of VFN-IF. We name this approach VFN-IFE, and the process can be represented as $f_{\text{ESM}} \circ \displaystyle f_{\text{if}}(\gT): \gT \rightarrow \vc$, where $\displaystyle f_{\text{if}}$ represents VFN-IF, and $f_{\text{ESM}}: \vc^{\text{o}} \rightarrow \vc$ denotes the ESM model fine-tuned with LoRA to correct the VFN-IF predictions $\vc^{\text{o}}$. For more VFN-IFE details, please refer to appendix~\ref{sec:vfn_ife}.

\section{Experiments}
As mentioned, we comprehensively validated the superiority of VFN-IF and VFN-Diff through experiments involving protein diffusion and inverse folding. The following sections will present the main experimental results for these two tasks separately. Additionally, we have included experiment details, extensive visual analyses and ablation studies based on inverse folding in the appendix~\ref{sec:supp_experiment}. We encourage readers to refer to the appendix for more in-depth information.
\begin{table}[]
\vspace{-1.8em}
\centering
\setlength\tabcolsep{10pt}
\resizebox{0.8\columnwidth}{!}{%
\begin{tabular}{c l ccc ccc}
\toprule
&\multirow{2}{*}{\raggedleft Model}      & \multicolumn{3}{c}{\textbf{Perplexity$\downarrow$}}   &\multicolumn{3}{c}{\textbf{Recovery$($\%$)\uparrow$}}         \\ \cmidrule(lr){3-5} \cmidrule(lr){6-8}
&                                        & \small{Short}         & \small{Single}& \small{All}& \small{Short}          & \small{Single}     & \small{All}      \\ \hline \addlinespace
\multirow{11}{*}{\rotatebox{90}{w/o ESM}} &StructGNN                               & 8.29          & 8.74          & 6.40          & 29.44          & 28.26            & 35.91     \\
&GraphTrans                              & 8.39          & 8.83          & 6.63          & 28.14          & 28.46            & 35.82 \\
&GCA                                     & 7.09          & 7.49          & 6.05          & 32.62          & 31.10            & 37.64 \\
&GVP                                     & 7.23          & 7.84          & 5.36          & 30.60          & 28.95            & 39.47 \\ 
&$\text{GVP-large}^\dag$                        & 7.68          & 6.12          & 6.17          & 32.60          & 39.40            & 39.20 \\
&AlphaDesign                             & 7.32          & 7.63          & 6.30          & 34.16          & 32.66            & 41.31 \\
&$\text{ESM-IF}^\dag$                           & 8.18          & 6.33          & 6.44          & 31.30          & 38.50            & 38.30 \\ 
&ProteinMPNN                             & 6.21          & 6.68          & 4.61          & 36.35          & 34.43            & 45.96 \\
&PiFold                                  & 6.04          & 6.31          & 4.55          & 39.84          & 38.53            & 51.66 \\
& \cellcolor{lm_purple}VFN-IF                                    &\cellcolor{lm_purple}{5.70}     & \cellcolor{lm_purple}{5.86}     &\cellcolor{lm_purple}{4.17}      & \cellcolor{lm_purple}\textbf{41.34} & \cellcolor{lm_purple}{40.98}       & \cellcolor{lm_purple}{ 54.74}      \\
& \cellcolor{lm_purple}VFN-IF+                       & \cellcolor{lm_purple}\textbf{5.31} & \cellcolor{lm_purple}\textbf{5.32} & \cellcolor{lm_purple}\textbf{3.66} & \cellcolor{lm_purple}{42.76 }    & \cellcolor{lm_purple}\textbf{41.77}   & \cellcolor{lm_purple}\textbf{57.14}\\ \hline \addlinespace
\multirow{3}{*}{\rotatebox{90}{ESM}} &$\text{ESM-IF}^\dag$                                    &6.05     & \textbf{4.00}     &4.01     & 38.10 & 51.50       & 51.60   \\
&LM-Design                                   &6.77     & 6.46     &4.52      & 37.88 & 42.47       & 55.65  \\
& \cellcolor{lm_purple}VFN-IFE                                    &\cellcolor{lm_purple}\textbf{4.92}     & \cellcolor{lm_purple}{4.22}     &\cellcolor{lm_purple}{\textbf{3.36}}      & \cellcolor{lm_purple}\textbf{50.00} & \cellcolor{lm_purple}{\textbf{52.13}}       & \cellcolor{lm_purple}{ \textbf{62.67}}    \\
\bottomrule
\end{tabular}
}
\vspace{-0.5em}
\caption{Experimental results comparison on the CATH dataset (inverse folding). Some results are reproduced by        \citet{gao2022pifold}. ``$\dag$'' denotes that the version of CATH used is 4.3, while for the remaining methods, the CATH version is 4.2.}\label{tab:cath}
\vspace{-2.0em}
\end{table}

\subsection{Inverse Folding}
In inverse folding, we followed the settings of \citep{gao2022pifold} and tested the sequence recovery performance of VFN on the CATH 4.2        \citep{orengo1997cath}, TS50, and TS500 datasets        \citep{li2014direct}. Furthermore, we tested the structure recovery performance of VFN and compared it with state-of-the-art models. Additionally, we demonstrated the superiority of VFN in trades-off between speed and accuracy.

In this task, we have devised three variations of the VFN model, namely VFN-IF, VFN-IF+, and VFN-IFE. VFN-IF represents the vanilla version of VFN, trained solely on the CATH 4.2 dataset. VFN-IF+ is an extended version of VFN-IF, scaled up to incorporate the entire PDB dataset during training. VFN-IFE denotes the version of VFN-IF equipped with an external knowledge base (ESM).

\begin{wraptable}{r}{0.58\columnwidth}
\vspace{0.2em}
\centering
\resizebox{0.58\columnwidth}{!}{
\begin{tabular}{l cc cc}
\toprule
\multicolumn{1}{l}{\multirow{2}{*}{Metric}} & \multicolumn{2}{c}{w/o ESM}      & \multicolumn{2}{c}{ESM}      \\ \cmidrule(l){2-3} \cmidrule(l){4-5} 
\multicolumn{1}{c}{}                       & PiFold    & \cellcolor{lm_purple}VFN-IF       & LM-Design    & \cellcolor{lm_purple}VFN-IFE       \\ \midrule
$\text{scTM}>0.5$                                   & 90.98\%           & \cellcolor{lm_purple}\textbf{92.37\%}          & 89.42\%          & \cellcolor{lm_purple}\textbf{93.29\%}          \\
$\text{scRMSD}<2$                                   & 60.35\%           & \cellcolor{lm_purple}\textbf{62.89\%}           & 58.41\%          & \cellcolor{lm_purple}\textbf{64.16\%}          \\
\bottomrule
\end{tabular}%
}
\vspace{-0.8em}
\caption{Experimental results on structure recovery (inverse folding). `$\text{scTM}>0.5$' represents the percentage of designed proteins that exhibit a structural similarity exceeding 0.5 with the desired protein. The same applies to '$\text{scRMSD}<2$'.}\label{tab:if_structure}
\vspace{-1.3em}
\end{wraptable}
\textbf{Sequence recovery.} The performance of VFN on the CATH dataset and TS50, TS500 is presented in Table~\ref{tab:cath}, Table~\ref{tab:ts50} and figure~\ref{fig:if_acc_length}, respectively. In these experiments, VFN is compared to other advanced models, such as StructGNN        \citep{ingraham2019generative}, GraphTrans        \citep{ingraham2019generative}, GCA        \citep{tan2022generative}, GVP        \citep{jing2020learning}, AlphaDesign        \citep{gao2022alphadesign}, ESM-IF        \citep{hsu2022learning}, ProteinMPNN        \citep{dauparas2022robust}, PiFold        \citep{gao2022pifold}, LM-design        \citep{zheng2023lm_design}, in terms of perplexity and sequence recovery performance. Table~\ref{tab:cath} provides detailed results for various subsets. Among the subsets considered in our analysis, the ``Short" subset refers to proteins with a length of up to 100 amino acids, while the ``Single" subset exclusively includes single chain proteins. `w/o ESM' and `ESM' refer to methods without and with the use of an external knowledge base (ESM), respectively.



\begin{wraptable}{r}{0.58\columnwidth}
\vspace{-1.0em}
\centering
\resizebox{0.58\columnwidth}{!}{
\begin{tabular}{c l cc cc}
\toprule
&\multicolumn{1}{l}{\multirow{2}{*}{Model}} & \multicolumn{2}{c}{\textbf{TS50}}      & \multicolumn{2}{c}{\textbf{TS500}}      \\ \cmidrule(l){3-4} \cmidrule(l){5-6} 
&\multicolumn{1}{c}{}                       & Perp.$\downarrow$    & Rec.($\%$)$\uparrow$       & Perp.$\downarrow$    & Rec.($\%$)$\uparrow$       \\ \midrule
\multirow{8}{*}{\rotatebox{90}{w/o ESM}}    &StructGNN                                   & 5.40           & 43.89          & 4.98          & 45.69          \\
&GraphTrans                                  & 5.60           & 42.20           & 5.16          & 44.66          \\
&GVP                                         & 4.71          & 44.14          & 4.20           & 49.14          \\
&GCA                                         & 5.09          & 47.02          & 4.72          & 47.74          \\
&AlphaDesign                                 & 5.25          & 48.36          & 4.93          & 49.23          \\
&ProteinMPNN                                 & 3.93          & 54.43          & 3.53          & 58.08          \\
&PiFold                                      & { 3.86}    & { 58.72}    & { 3.44}    & { 60.42}    \\
&\cellcolor{lm_purple}VFN-IF                                      & \cellcolor{lm_purple}\textbf{3.58} & \cellcolor{lm_purple}\textbf{59.54} & \cellcolor{lm_purple}\textbf{3.19} & \cellcolor{lm_purple}\textbf{63.65} \\ \hline \addlinespace
\multirow{2}{*}{\rotatebox{90}{ESM}}&LM-Design                                 & 3.50          & 57.89          & 3.19          & 67.78          \\
&\cellcolor{lm_purple}VFN-IFE                                      & \cellcolor{lm_purple}\textbf{2.52} & \cellcolor{lm_purple}\textbf{73.30} & \cellcolor{lm_purple}\textbf{2.54} & \cellcolor{lm_purple}\textbf{72.49} \\

\bottomrule
\end{tabular}%
}
\vspace{-0.8em}
\caption{Experimental results on TS50 and TS500 (inverse folding). }\label{tab:ts50}
\vspace{-1.6em}
\end{wraptable}


\textbf{Structure recovery.}
We compared VFN's performance in terms of protein structure recovery with several advanced methods on CATH 4.2, as shown in table~\ref{tab:if_structure}. We followed standard evaluation procedures\citep{yim2023se}. Specifically, we used ESMFold to predict whether sequences designed by inverse folding networks could fold proteins into the desired structures, i.e., the input structures of the inverse folding network. We employed two metrics, scTM$\uparrow$ and scRMSD$\downarrow$, to assess the similarity between the desired protein structures and the structures of proteins designed through inverse folding. Our experimental results demonstrate that VFN has a significant advantage compared to state-of-the-art methods.

\textbf{Speed and accuracy trade-off.} In terms of the trade-off between speed and accuracy, we compared VFN-IF with the SoTA model in this regard, PiFold. We conducted comparisons between different layers of VFN and PiFold, as shown in figure~\ref{fig:first_page} (A). Experimental results demonstrate that VFN-IF achieves SoTA efficiency. Even with just 5 layers, VFN-IF achieves higher accuracy (52.68\%)  than a 10-layer PiFold while maintaining faster inference speeds. Furthermore, PiFold's accuracy saturates at 10 layers, whereas VFN-IF does not encounter this issue.
For more details, please refer to table~ \ref{tab:layers} in the appendix.

\subsection{Protein Diffusion}
We followed the settings and benchmarks of FrameDiff        \citep{yim2023se}, conducting a detailed comparison between VFN-Diff and FrameDiff in terms of designability and diversity. It is worth emphasizing that FrameDiff and VFN-Diff differ only in the protein structure encoder (VFN \textit{vs.} IPA). All other settings are identical, and the parameter counts are similar (18.3M \textit{vs.} 17.4M), making it an ablation study. On the other hand, RFDiffusion        \citep{watson2023novo} is a recent advance in the field. However, as FrameDiff pointed out, RFDiffusion performs noticeably worse than FrameDiff in the same setting (without pre-trained weights). Additionally, RFDiffusion has a larger number of parameters and is trained on larger datasets (i.e., complex data). Therefore, comparing with RFDiffusion is beyond the scope\footnote{Our scope is to overcome the atom representation bottleneck using VFN.} of this work. We leave these engineering implementations for future research.
\begin{wrapfigure}{r}{0.50\columnwidth}
    \vspace{-0.0em}
    \includegraphics[width=0.50\columnwidth]{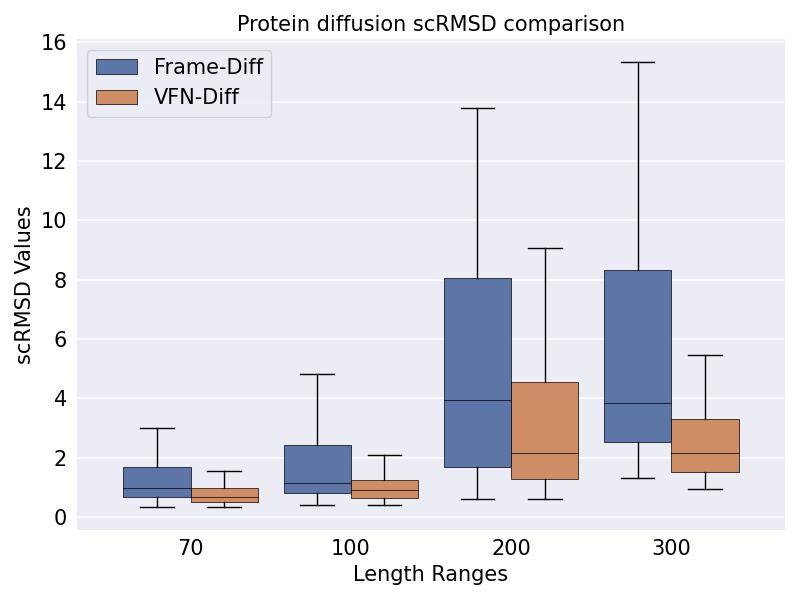}
    \vspace{-1.0em}
    \caption{Visualization of designability. $N_{\text{seq}}$ represents the number of attempts when reconstructing protein structures using Protein MPNN and ESMFold.}\label{fig:diff_scRMSD}
    \vspace{-2.0em}
\end{wrapfigure}

\textbf{Designability.} As shown in table~\ref{tab:diversity_disgnability} and figure~\ref{fig:diff_scRMSD}, we compare the protein designability of VFN-Diff with that of FrameDiff. For evaluation, Proteins generated by the diffusion model are reconstructed using the inverse folding network (ProteinMPNN) and structural prediction network (ESMFold). The designability of the generated proteins is then assessed by comparing their structural similarity (scTM, scRMSD) to the reconstructed proteins. Experimental results demonstrate that VFN-Diff significantly outperforms FrameDiff in terms of designability.


\begin{wraptable}{r}{0.7\columnwidth}
\vspace{-1.0em}
\centering
\resizebox{0.7\columnwidth}{!}{
\begin{tabular}{c l l ccccc}
\toprule
                                                            &\multirow{3}{*}{\diagbox[trim=l,height=3.5\line]{Metric}{Setting}}    & Noise Scale                           & 1.0  & 0.5 & 0.1 & 0.1 & 0.1    \\ \cmidrule(l){3-8}
                                                            &                           & Num. Step                             & 500  & 500 & 500 & 500 & 100    \\ \cmidrule(l){3-8}
                                                            &                           & Num. Seq.                             & 8  & 8 & 8 & 100 & 8    \\ \midrule \addlinespace
\multirow{4}{*}{\rotatebox{90}{\small{Designability}}}      &\multirow{2}{*}{$\text{scTM}_{0.5}\uparrow$}    & FrameDiff        & 53.58\% & 76.42\% & 77.41\% & 87.04\% & 76.67\%   \\
                                                            &                           & \cellcolor{lm_purple}VFN-Diff         & \cellcolor{lm_purple}\textbf{67.04\%}  & \cellcolor{lm_purple}\textbf{81.23\%} & \cellcolor{lm_purple}\textbf{83.95\%} & \cellcolor{lm_purple}\textbf{92.84\%} & \cellcolor{lm_purple}\textbf{83.83\%}          \\ \cmidrule(l){2-8}
                                                            &\multirow{2}{*}{$\text{scRMSD}_2\downarrow$}& FrameDiff            & 10.62\% & 23.46\% & 28.02\% & 37.78\% & 26.42\% \\
                                                            &                           & \cellcolor{lm_purple}VFN-Diff         & \cellcolor{lm_purple}\textbf{25.93\%} & \cellcolor{lm_purple}\textbf{40.00\%} & \cellcolor{lm_purple}\textbf{44.20\%} & \cellcolor{lm_purple}\textbf{56.30\%} & \cellcolor{lm_purple}\textbf{40.25\%}    \\ \midrule \addlinespace
\multirow{4}{*}{\rotatebox{90}{\small{Diversity}}}          &\multirow{2}{*}{Diversity  $\uparrow$}& FrameDiff                             & 51.98\%  & 74.57\% & 75.56\% & 85.43\% & 74.94\%          \\
                                                            &                           & \cellcolor{lm_purple}VFN-Diff         & \cellcolor{lm_purple}\textbf{66.54\%}  & \cellcolor{lm_purple}\textbf{80.49\%} & \cellcolor{lm_purple}\textbf{83.33\%} & \cellcolor{lm_purple}\textbf{90.61\%} & \cellcolor{lm_purple}\textbf{82.59\%}          \\ \cmidrule(l){2-8}
                                                            &\multirow{2}{*}{$\text{pdbTM}_{0.7}\downarrow$}& FrameDiff                             & 5  & 30 & 37 & 86 & 35          \\
                                                            &                           & \cellcolor{lm_purple}VFN-Diff         & \cellcolor{lm_purple}\textbf{9}  & \cellcolor{lm_purple}\textbf{47} & \cellcolor{lm_purple}\textbf{54} & \cellcolor{lm_purple}\textbf{102} & \cellcolor{lm_purple}\textbf{48}          \\
\bottomrule
\end{tabular}%
}
\vspace{-0.8em}
\caption{Experimental results on protein structure diffusion assessing the designability and diversity of VFN-IF. `scTM$_{0.5}$' and `scRMSD$_2$' represent the percentages of generated proteins with scTM greater than 0.5 and scRMSD less than 2, respectively, assessing structural similarity. pdbTM$_{0.7}$' signifies the count of generated proteins with a pdbTM score less than 0.7, measuring their similarity to existing proteins in the PDB database. Further details on Diversity', `pdbTM', and the setting parameters can be found in Appendix~\ref{sec:supp_metric}.}\label{tab:diversity_disgnability}
\vspace{-3.0em}
\end{wraptable}


\textbf{Diversity.} Diversity is an important metric for generative models. Similarly, we followed FrameDiff's relevant evaluation metrics, namely `diversity' and `pdbTM', to compare the diversity of VFN-IF and FrameDiff. Specifically, `diversity' represents the clustering center density of generated samples. To be more specific, we first excluded undesignable proteins (scTM $< 0.5$). Then, we used MaxCluster        \citep{herbert2008maxcluster} to cluster the remaining samples and obtain clustering centers. Finally, clustering center density can be calculated as follows: (number of clustering center) / (number of generated samples). `pdbTM' represents the structural similarity of generated samples to the most similar structures in the PDB database.

\textbf{Visualization Comparison.} Through visualization, we observed that FrameDiff suffers from very low designability when designing longer protein sequences, while VFN-Diff does not have this issue. We selected typical generated protein sequences and compared the results generated by VFN-Diff and FrameDiff through visualization. The visual results demonstrate the superiority of VFN in encoding IPA on the encoder, as shown in figure~\ref{fig:diffusion_vis}.



\vspace{1.0em}
\section{Conclusion}
By introducing VFN, we address the atom representation bottleneck though the vector field operator, enhancing its capacity for modeling frames. We demonstrate VFN's expressiveness through comprehensive experiments in \textit{de novo} protein design. Significant improvements or state-of-the-art performance are achieved in protein diffusion and inverse folding tasks.

\bibliography{iclr2024_conference}

\begin{thebibliography}{51}
\providecommand{\natexlab}[1]{#1}
\providecommand{\url}[1]{\texttt{#1}}
\expandafter\ifx\csname urlstyle\endcsname\relax
  \providecommand{\doi}[1]{doi: #1}\else
  \providecommand{\doi}{doi: \begingroup \urlstyle{rm}\Url}\fi

\bibitem[Anishchenko et~al.(2021)Anishchenko, Pellock, Chidyausiku, Ramelot, Ovchinnikov, Hao, Bafna, Norn, Kang, and Bera]{anishchenko2021novo}
Ivan Anishchenko, Samuel~J. Pellock, Tamuka~M. Chidyausiku, Theresa~A. Ramelot, Sergey Ovchinnikov, Jingzhou Hao, Khushboo Bafna, Christoffer Norn, Alex Kang, and et~al. Bera, Asim~K.
\newblock De novo protein design by deep network hallucination.
\newblock \emph{Nature}, 600\penalty0 (7889):\penalty0 547--552, 2021.

\bibitem[Baek et~al.(2021)Baek, DiMaio, Anishchenko, Dauparas, Ovchinnikov, Lee, Wang, Cong, Kinch, and Schaeffer]{baek2021accurate}
Minkyung Baek, Frank DiMaio, Ivan Anishchenko, Justas Dauparas, Sergey Ovchinnikov, Gyu~Rie Lee, Jue Wang, Qian Cong, Lisa~N. Kinch, and et~al. Schaeffer, R.~Dustin.
\newblock Accurate prediction of protein structures and interactions using a three-track neural network.
\newblock \emph{Science}, 373\penalty0 (6557):\penalty0 871--876, 2021.

\bibitem[Baldassarre et~al.(2021)Baldassarre, Men{\'e}ndez~Hurtado, Elofsson, and Azizpour]{baldassarre2021graphqa}
Federico Baldassarre, David Men{\'e}ndez~Hurtado, Arne Elofsson, and Hossein Azizpour.
\newblock Graphqa: protein model quality assessment using graph convolutional networks.
\newblock \emph{Bioinformatics}, 37\penalty0 (3):\penalty0 360--366, 2021.

\bibitem[Berman et~al.(2000)Berman, Westbrook, Feng, Gilliland, Bhat, Weissig, Shindyalov, and Bourne]{berman2000pdb}
Helen~M. Berman, John Westbrook, Zukang Feng, Gary Gilliland, T.~N. Bhat, Helge Weissig, Ilya~N. Shindyalov, and Philip~E. Bourne.
\newblock {The Protein Data Bank}.
\newblock \emph{Nucleic Acids Research}, 28\penalty0 (1):\penalty0 235--242, 01 2000.
\newblock ISSN 0305-1048.

\bibitem[Burley et~al.(2017)Burley, Berman, Kleywegt, Markley, Nakamura, and Velankar]{burley2017protein}
Stephen~K. Burley, Helen~M. Berman, Gerard~J. Kleywegt, John~L. Markley, Haruki Nakamura, and Sameer Velankar.
\newblock Protein data bank (pdb): the single global macromolecular structure archive.
\newblock \emph{Protein Crystallography: Methods and Protocols}, pp.\  627--641, 2017.

\bibitem[Cao et~al.(2021)Cao, Das, Chenthamarakshan, Chen, Melnyk, and Shen]{cao2021fold2seq}
Yue Cao, Payel Das, Vijil Chenthamarakshan, Pin-Yu Chen, Igor Melnyk, and Yang Shen.
\newblock Fold2seq: A joint sequence (1d)-fold (3d) embedding-based generative model for protein design.
\newblock In \emph{International Conference on Machine Learning}, pp.\  1261--1271. PMLR, 2021.

\bibitem[Dauparas et~al.(2022)Dauparas, Anishchenko, Bennett, Bai, Ragotte, Milles, Wicky, Courbet, de~Haas, and Bethel]{dauparas2022robust}
Justas Dauparas, Ivan Anishchenko, Nathaniel Bennett, Hua Bai, Robert~J. Ragotte, Lukas~F. Milles, Basile~IM Wicky, Alexis Courbet, Rob~J. de~Haas, and et~al. Bethel, Neville.
\newblock Robust deep learning--based protein sequence design using proteinmpnn.
\newblock \emph{Science}, 378\penalty0 (6615):\penalty0 49--56, 2022.

\bibitem[Derevyanko et~al.(2018)Derevyanko, Grudinin, Bengio, and Lamoureux]{derevyanko2018deep}
Georgy Derevyanko, Sergei Grudinin, Yoshua Bengio, and Guillaume Lamoureux.
\newblock Deep convolutional networks for quality assessment of protein folds.
\newblock \emph{Bioinformatics}, 34\penalty0 (23):\penalty0 4046--4053, 2018.

\bibitem[Dumortier et~al.(2022)Dumortier, Liutkus, Carr{\'e}, and Krouk]{dumortier2022petribert}
Baldwin Dumortier, Antoine Liutkus, Cl{\'e}ment Carr{\'e}, and Gabriel Krouk.
\newblock Petribert: Augmenting bert with tridimensional encoding for inverse protein folding and design.
\newblock \emph{BioRxiv}, pp.\  2022--08, 2022.

\bibitem[Fu et~al.(2023)Fu, Yan, Wang, Au, McThrow, Komikado, Maruhashi, Uchino, Qian, and Ji]{fu2023latent}
Cong Fu, Keqiang Yan, Limei Wang, Wing~Yee Au, Michael McThrow, Tao Komikado, Koji Maruhashi, Kanji Uchino, Xiaoning Qian, and Shuiwang Ji.
\newblock A latent diffusion model for protein structure generation, 2023.

\bibitem[Fuchs et~al.(2020)Fuchs, Worrall, Fischer, and Welling]{fuchs2020se}
Fabian Fuchs, Daniel Worrall, Volker Fischer, and Max Welling.
\newblock {SE}(3)-transformers: 3d roto-translation equivariant attention networks.
\newblock \emph{Advances in Neural Information Processing Systems}, 33:\penalty0 1970--1981, 2020.

\bibitem[Gao et~al.(2022{\natexlab{a}})Gao, Tan, and Li]{gao2022pifold}
Zhangyang Gao, Cheng Tan, and Stan Li.
\newblock {PiFold}: Toward effective and efficient protein inverse folding.
\newblock \emph{International Conference on Learning Representations}, 2022{\natexlab{a}}.

\bibitem[Gao et~al.(2022{\natexlab{b}})Gao, Tan, and Li]{gao2022alphadesign}
Zhangyang Gao, Cheng Tan, and Stan~Z Li.
\newblock Alphadesign: A graph protein design method and benchmark on alphafolddb.
\newblock \emph{arXiv preprint arXiv:2202.01079}, 2022{\natexlab{b}}.

\bibitem[Gao et~al.(2023{\natexlab{a}})Gao, Tan, and Li]{gao2023diffsds}
Zhangyang Gao, Cheng Tan, and Stan~Z. Li.
\newblock Diffsds: A language diffusion model for protein backbone inpainting under geometric conditions and constraints, 2023{\natexlab{a}}.

\bibitem[Gao et~al.(2023{\natexlab{b}})Gao, Tan, and Li]{gao2023knowledge}
Zhangyang Gao, Cheng Tan, and Stan~Z Li.
\newblock Knowledge-design: Pushing the limit of protein deign via knowledge refinement.
\newblock \emph{arXiv preprint arXiv:2305.15151}, 2023{\natexlab{b}}.

\bibitem[Herbert \& Sternberg(2008)Herbert and Sternberg]{herbert2008maxcluster}
Alex Herbert and MJE Sternberg.
\newblock Maxcluster: a tool for protein structure comparison and clustering, 2008.

\bibitem[Hermosilla \& Ropinski(2022)Hermosilla and Ropinski]{hermosilla2022contrastive}
Pedro Hermosilla and Timo Ropinski.
\newblock Contrastive representation learning for 3d protein structures.
\newblock \emph{arXiv preprint arXiv:2205.15675}, 2022.

\bibitem[Hermosilla et~al.(2020)Hermosilla, Sch{\"a}fer, Lang, Fackelmann, V{\'a}zquez, Kozlikova, Krone, Ritschel, and Ropinski]{hermosilla2020intrinsic}
Pedro Hermosilla, Marco Sch{\"a}fer, Matej Lang, Gloria Fackelmann, Pere-Pau V{\'a}zquez, Barbora Kozlikova, Michael Krone, Tobias Ritschel, and Timo Ropinski.
\newblock Intrinsic-extrinsic convolution and pooling for learning on 3d protein structures.
\newblock In \emph{International Conference on Learning Representations}, 2020.

\bibitem[Hie et~al.(2022)Hie, Candido, Lin, Kabeli, Rao, Smetanin, Sercu, and Rives]{brain2022esmfold}
Brian Hie, Salvatore Candido, Zeming Lin, Ori Kabeli, Roshan Rao, Nikita Smetanin, Tom Sercu, and Alexander Rives.
\newblock A high-level programming language for generative protein design.
\newblock \emph{bioRxiv}, 2022.

\bibitem[Hou et~al.(2018)Hou, Adhikari, and Cheng]{hou2018deepsf}
Jie Hou, Badri Adhikari, and Jianlin Cheng.
\newblock Deepsf: deep convolutional neural network for mapping protein sequences to folds.
\newblock \emph{Bioinformatics}, 34\penalty0 (8):\penalty0 1295--1303, 2018.

\bibitem[Hsu et~al.(2022)Hsu, Verkuil, Liu, Lin, Hie, Sercu, Lerer, and Rives]{hsu2022learning}
Chloe Hsu, Robert Verkuil, Jason Liu, Zeming Lin, Brian Hie, Tom Sercu, Adam Lerer, and Alexander Rives.
\newblock Learning inverse folding from millions of predicted structures.
\newblock In \emph{International Conference on Machine Learning}, pp.\  8946--8970. PMLR, 2022.

\bibitem[Hu et~al.(2021)Hu, Shen, Wallis, Allen-Zhu, Li, Wang, Wang, and Chen]{hu2021lora}
Edward~J Hu, Yelong Shen, Phillip Wallis, Zeyuan Allen-Zhu, Yuanzhi Li, Shean Wang, Lu~Wang, and Weizhu Chen.
\newblock Lora: Low-rank adaptation of large language models.
\newblock \emph{arXiv preprint arXiv:2106.09685}, 2021.

\bibitem[Huang et~al.(2022)Huang, Xu, Hu, Liu, Liao, Zhang, Huang, Hong, Chen, and Liu]{huang2022backbone}
Bin Huang, Yang Xu, Xiuhong Hu, Yongrui Liu, Shanhui Liao, Jiahai Zhang, Chengdong Huang, Jingjun Hong, Quan Chen, and Haiyan Liu.
\newblock A backbone-centred energy function of neural networks for protein design.
\newblock \emph{Nature}, 602\penalty0 (7897):\penalty0 523--528, 2022.

\bibitem[Huang et~al.(2016)Huang, Boyken, and Baker]{huang2016coming}
Po-Ssu Huang, Scott~E Boyken, and David Baker.
\newblock The coming of age of de novo protein design.
\newblock \emph{Nature}, 537\penalty0 (7620):\penalty0 320--327, 2016.

\bibitem[Ingraham et~al.(2019)Ingraham, Garg, Barzilay, and Jaakkola]{ingraham2019generative}
John Ingraham, Vikas Garg, Regina Barzilay, and Tommi Jaakkola.
\newblock Generative models for graph-based protein design.
\newblock \emph{Advances in Neural Information Processing Systems}, 32, 2019.

\bibitem[Jendrusch et~al.(2021)Jendrusch, Korbel, and Sadiq]{jendrusch2021alphadesign}
Michael Jendrusch, Jan Korbel, and Kashif Sadiq.
\newblock Alphadesign: A de novo protein design framework based on {AlphaFold}.
\newblock \emph{BioRxiv}, pp.\  2021--10, 2021.

\bibitem[Jin et~al.(2022)Jin, Barzilay, and Jaakkola]{jin2022antibody}
Wengong Jin, Regina Barzilay, and Tommi Jaakkola.
\newblock Antibody-antigen docking and design via hierarchical equivariant refinement.
\newblock \emph{arXiv preprint arXiv:2207.06616}, 2022.

\bibitem[Jing et~al.(2020)Jing, Eismann, Suriana, Townshend, and Dror]{jing2020learning}
Bowen Jing, Stephan Eismann, Patricia Suriana, Raphael Townshend, and Ron Dror.
\newblock Learning from protein structure with geometric vector perceptrons.
\newblock \emph{International Conference on Learning Representations}, 2020.

\bibitem[Jumper et~al.(2021)Jumper, Evans, Pritzel, Green, Figurnov, Ronneberger, Tunyasuvunakool, Bates, {\v{Z}}{\'\i}dek, and Potapenko]{jumper2021highly}
John Jumper, Richard Evans, Alexander Pritzel, Tim Green, Michael Figurnov, Olaf Ronneberger, Kathryn Tunyasuvunakool, Russ Bates, Augustin {\v{Z}}{\'\i}dek, and et~al. Potapenko, Anna.
\newblock Highly accurate protein structure prediction with alphafold.
\newblock \emph{Nature}, 596\penalty0 (7873):\penalty0 583--589, 2021.

\bibitem[Karimi et~al.(2020)Karimi, Zhu, Cao, and Shen]{karimi2020novo}
Mostafa Karimi, Shaowen Zhu, Yue Cao, and Yang Shen.
\newblock De novo protein design for novel folds using guided conditional wasserstein generative adversarial networks.
\newblock \emph{Journal of Chemical Information and Modeling}, 60\penalty0 (12):\penalty0 5667--5681, 2020.

\bibitem[Li et~al.(2022)Li, Luo, Deng, Cheng, Guan, Guibas, Peng, and Ma]{li2022directed}
Jiahan Li, Shitong Luo, Congyue Deng, Chaoran Cheng, Jiaqi Guan, Leonidas Guibas, Jian Peng, and Jianzhu Ma.
\newblock Directed weight neural networks for protein structure representation learning.
\newblock \emph{arXiv preprint arXiv:2201.13299}, 2022.

\bibitem[Li et~al.(2014)Li, Yang, Faraggi, Zhan, and Zhou]{li2014direct}
Zhixiu Li, Yuedong Yang, Eshel Faraggi, Jian Zhan, and Yaoqi Zhou.
\newblock Direct prediction of profiles of sequences compatible with a protein structure by neural networks with fragment-based local and energy-based nonlocal profiles.
\newblock \emph{Proteins: Structure, Function, and Bioinformatics}, 82\penalty0 (10):\penalty0 2565--2573, 2014.

\bibitem[Lin et~al.(2023)Lin, Akin, Rao, Hie, Zhu, Lu, Smetanin, dos Santos~Costa, Fazel-Zarandi, Sercu, Candido, and Rives]{lin2022language}
Zeming Lin, Halil Akin, Roshan Rao, Brian Hie, Zhongkai Zhu, Wenting Lu, Nikita Smetanin, Allan dos Santos~Costa, Maryam Fazel-Zarandi, Tom Sercu, Sal Candido, and Alexander Rives.
\newblock Evolutionary-scale prediction of atomic-level protein structure with a language model.
\newblock \emph{Science}, 2023.

\bibitem[Liu \& Kuhlman(2006)Liu and Kuhlman]{liu2006rosettadesign}
Yi~Liu and Brian Kuhlman.
\newblock Rosettadesign server for protein design.
\newblock \emph{Nucleic acids research}, 34\penalty0 (suppl\_2):\penalty0 W235--W238, 2006.

\bibitem[McPartlon et~al.(2022)McPartlon, Lai, and Xu]{mcpartlon2022deep}
Matt McPartlon, Ben Lai, and Jinbo Xu.
\newblock A deep {SE}(3)-equivariant model for learning inverse protein folding.
\newblock \emph{BioRxiv}, pp.\  2022--04, 2022.

\bibitem[Orengo et~al.(1997)Orengo, Michie, Jones, Jones, Swindells, and Thornton]{orengo1997cath}
Christine Orengo, Alex Michie, Susan Jones, David Jones, Mark Swindells, and Janet Thornton.
\newblock Cath--a hierarchic classification of protein domain structures.
\newblock \emph{Structure}, 5\penalty0 (8):\penalty0 1093--1109, 1997.

\bibitem[Ovchinnikov \& Huang(2021)Ovchinnikov and Huang]{ovchinnikov2021structure}
Sergey Ovchinnikov and Po-Ssu Huang.
\newblock Structure-based protein design with deep learning.
\newblock \emph{Current Opinion in Structural Biology}, 65:\penalty0 136--144, 2021.

\bibitem[Shroff et~al.(2019)Shroff, Cole, Morrow, Diaz, Donnell, Gollihar, Ellington, and Thyer]{shroff2019structure}
Raghav Shroff, Austin Cole, Barrett Morrow, Daniel Diaz, Isaac Donnell, Jimmy Gollihar, Andrew Ellington, and Ross Thyer.
\newblock A structure-based deep learning framework for protein engineering.
\newblock \emph{BioRxiv}, 2019.

\bibitem[Tan et~al.(2022)Tan, Gao, Xia, and Li]{tan2022generative}
Cheng Tan, Zhangyang Gao, Jun Xia, and Stan Li.
\newblock Generative de novo protein design with global context.
\newblock \emph{arXiv preprint arXiv:2204.10673}, 2022.

\bibitem[Townshend et~al.(2021)Townshend, V{\"o}gele, Suriana, Derry, Powers, Laloudakis, Balachandar, Jing, Anderson, Eismann, et~al.]{townshend2021atom3d}
Raphael John~Lamarre Townshend, Martin V{\"o}gele, Patricia~Adriana Suriana, Alexander Derry, Alexander Powers, Yianni Laloudakis, Sidhika Balachandar, Bowen Jing, Brandon~M Anderson, Stephan Eismann, et~al.
\newblock Atom3d: Tasks on molecules in three dimensions.
\newblock In \emph{Thirty-fifth Conference on Neural Information Processing Systems Datasets and Benchmarks Track (Round 1)}, 2021.

\bibitem[van Kempen et~al.(2023)van Kempen, Kim, Tumescheit, Mirdita, Lee, Gilchrist, S{\"o}ding, and Steinegger]{van2023foldseek}
Michel van Kempen, Stephanie~S. Kim, Charlotte Tumescheit, Milot Mirdita, Jeongjae Lee, Cameron~L.M. Gilchrist, Johannes S{\"o}ding, and Martin Steinegger.
\newblock Fast and accurate protein structure search with foldseek.
\newblock \emph{bioRxiv}, 2023.

\bibitem[Veli{\v{c}}kovi{\'c} et~al.(2017)Veli{\v{c}}kovi{\'c}, Cucurull, Casanova, Romero, Lio, and Bengio]{velivckovic2017graph}
Petar Veli{\v{c}}kovi{\'c}, Guillem Cucurull, Arantxa Casanova, Adriana Romero, Pietro Lio, and Yoshua Bengio.
\newblock Graph attention networks.
\newblock \emph{arXiv preprint arXiv:1710.10903}, 2017.

\bibitem[Wang et~al.(2022{\natexlab{a}})Wang, Liu, Liu, Kurtin, and Ji]{wang2022learning}
Limei Wang, Haoran Liu, Yi~Liu, Jerry Kurtin, and Shuiwang Ji.
\newblock Learning hierarchical protein representations via complete 3d graph networks.
\newblock In \emph{The Eleventh International Conference on Learning Representations}, 2022{\natexlab{a}}.

\bibitem[Wang et~al.(2022{\natexlab{b}})Wang, Liu, Lin, Liu, and Ji]{wang2022comenet}
Limei Wang, Yi~Liu, Yuchao Lin, Haoran Liu, and Shuiwang Ji.
\newblock Comenet: Towards complete and efficient message passing for 3d molecular graphs.
\newblock \emph{Advances in Neural Information Processing Systems}, 35:\penalty0 650--664, 2022{\natexlab{b}}.

\bibitem[Watson et~al.(2023)Watson, Juergens, Bennett, Trippe, Yim, Eisenach, Ahern, Borst, Ragotte, Milles, et~al.]{watson2023novo}
Joseph~L Watson, David Juergens, Nathaniel~R Bennett, Brian~L Trippe, Jason Yim, Helen~E Eisenach, Woody Ahern, Andrew~J Borst, Robert~J Ragotte, Lukas~F Milles, et~al.
\newblock De novo design of protein structure and function with rfdiffusion.
\newblock \emph{Nature}, pp.\  1--3, 2023.

\bibitem[Wu et~al.(2021)Wu, Johnston, Arnold, and Yang]{wu2021protein}
Zachary Wu, Kadina~E. Johnston, Frances~H. Arnold, and Kevin~K. Yang.
\newblock Protein sequence design with deep generative models.
\newblock \emph{Current Opinion in Structural Biology}, 65:\penalty0 18--27, 2021.

\bibitem[Yim et~al.(2023)Yim, Trippe, De~Bortoli, Mathieu, Doucet, Barzilay, and Jaakkola]{yim2023se}
Jason Yim, Brian~L Trippe, Valentin De~Bortoli, Emile Mathieu, Arnaud Doucet, Regina Barzilay, and Tommi Jaakkola.
\newblock Se (3) diffusion model with application to protein backbone generation.
\newblock \emph{arXiv preprint arXiv:2302.02277}, 2023.

\bibitem[Zhang et~al.(2020)Zhang, Chen, Wang, Lo, Liu, Wu, and Zhang]{zhang2020prodconn}
Yuan Zhang, Yang Chen, Chenran Wang, Chun-Chao Lo, Xiuwen Liu, Wei Wu, and Jinfeng Zhang.
\newblock Prodconn: Protein design using a convolutional neural network.
\newblock \emph{Proteins: Structure, Function, and Bioinformatics}, 88\penalty0 (7):\penalty0 819--829, 2020.

\bibitem[Zhang et~al.(2022)Zhang, Xu, Jamasb, Chenthamarakshan, Lozano, Das, and Tang]{zhang2022protein}
Zuobai Zhang, Minghao Xu, Arian~Rokkum Jamasb, Vijil Chenthamarakshan, Aurelie Lozano, Payel Das, and Jian Tang.
\newblock Protein representation learning by geometric structure pretraining.
\newblock In \emph{The Eleventh International Conference on Learning Representations}, 2022.

\bibitem[Zheng et~al.(2023)Zheng, Deng, Xue, Zhou, YE, and Gu]{zheng2023lm_design}
Zaixiang Zheng, Yifan Deng, Dongyu Xue, Yi~Zhou, Fei YE, and Quanquan Gu.
\newblock Structure-informed language models are protein designers.
\newblock In \emph{International Conference on Machine Learning}, 2023.

\bibitem[Zhou et~al.(2023)Zhou, Gao, Ding, Zheng, Xu, Wei, Zhang, and Ke]{zhou2023uni}
Gengmo Zhou, Zhifeng Gao, Qiankun Ding, Hang Zheng, Hongteng Xu, Zhewei Wei, Linfeng Zhang, and Guolin Ke.
\newblock Uni-mol: A universal 3d molecular representation learning framework.
\newblock 2023.

\end{thebibliography}
\bibliographystyle{iclr2024_conference}

\newpage
\appendix
\section{Appendix}


\subsection{Methodology}

\subsubsection{Multi-Layer Vector Perceptron Module}\label{sec:vmlp}
In this section, we elaborate on the process of the multi-layer vector perceptron module, $\operatorname{V-MLP}$, as illustrated in Algorithm~\ref{pseudocode:vmlp}:
\begin{algorithm}[H]
    \setstretch{1.5}
    \caption{The pseudo-code for the multi-layer vector perceptron module.}
    \label{pseudocode:vmlp}
    \begin{algorithmic}[1]
        \STATE def $\operatorname{V-MLP}$($\rmQ_i, \rmQ_i^\text{o}$): \\
        \STATE \qquad $\rvw^\text{c} = \{w^\text{c}_{k,l} \in \R \}_{k,l=1,...,d_q}, \rvw^\text{d} = \{w^\text{d}_{k,l} \in \R \}_{k,l=1,...,d_q}$ \\
        \qquad \textit{\textcolor[RGB]{136,16,16}{\#\ \ \ initialize learnable weight $\rvw^\text{c}, \rvw^\text{d}$}}
        \STATE \qquad $\vec{\rvv}_{k} = \sum_l w^\text{c}_{k,l} \vec{\rvq}_{l} + \sum_l w^\text{d}_{k,l} \vec{\rvq}_{l}^\text{~o}$ \hfill $\vec{\rvv}_{k} \in \R^3\ ,\vec{\rvq}_{l} \in \rmQ_i, \vec{\rvq}_{l}^\text{~o} \in \rmQ_i^\text{o}$\\
        \STATE \qquad $\vec{\vu}_{k} = \frac{{\vec{\rvw}_{k}} \cdot \vec{\rvv}_{k}}{\Vert {\vec{\rvw}_{k}}\Vert \Vert \vec{\rvv}_{k}\Vert}  \vec{\rvv}_{k}$ \hfill $\vec{\rvw}_{k} \in \R^3,\ \ ({\vec{\rvw}_{k}} \cdot \vec{\rvv}_{k})  \in \R$\\
        \qquad \textit{\textcolor[RGB]{136,16,16}{\#\ \ \ $\vec{\rvw}_{k}$ are learnable weights.}}\\
        \STATE \qquad $\rvw^\text{e} = \{w^\text{e}_{m,k} \in \R \}_{m,k=1,...,d_q}$ \\
        \qquad \textit{\textcolor[RGB]{136,16,16}{\#\ \ \ initialize learnable weight $\rvw^\text{e}$.}}\\
        \STATE \qquad $\vec{\ve}_{m} = \sum_k w^\text{e}_{m,k} \vec{\vu}_{k}$ \hfill $\vec{\ve}_{m} \in \R^3$\\
        \STATE \qquad return $\{\vec{\ve}_{m}\}_{m=1,....,d_q}$ \\
    \end{algorithmic}
\end{algorithm}

\subsubsection{Implementation of VFN-Diff}~\label{sec:vfn_diff}
\begin{figure}[h]
\centering
    \includegraphics[width=1.0\linewidth]{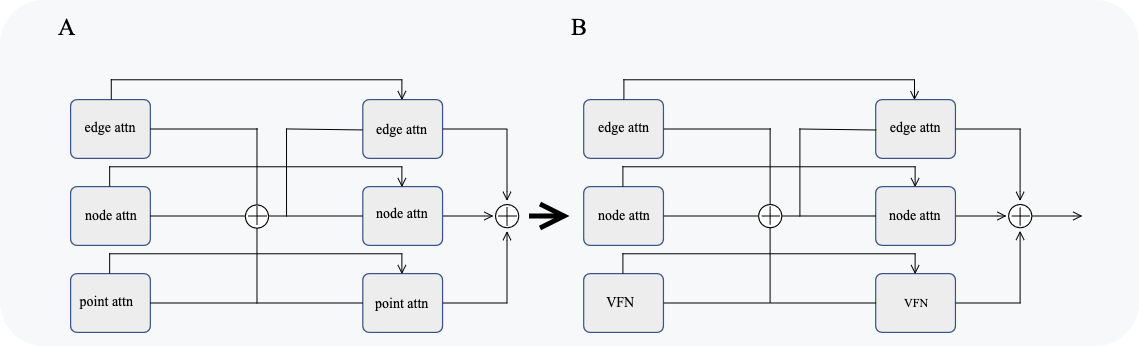}
    \caption{In the context of FrameDiff (A) and VFN-Diff (B), the traditional point attention module in FrameDiff has been replaced with a VFN layer, introducing enhanced geometric feature extraction capabilities}\label{fig:framediff vfndiff}
\end{figure}

We have implemented VFN-Diff based on FrameDiff, as shown in figure~\ref{fig:framediff vfndiff}. Apart from modifications to the structural encoder, IPA, all other parts remain unchanged. Within IPA, attention is divided into three components: node attention, edge attention, and point attention. In the case of VFN-Diff, we replace the point attention component in IPA with the VFN layer, while keeping the remaining portions unaltered. Despite the removal of node attention and edge attention in VFN-Diff, the model still functions effectively, achieving performance comparable to FrameDiff. However, it is important to note that, at this stage, VFN-Diff's parameter count is only one-ninth that of FrameDiff. Clearly, such a comparison would be unfair. Therefore, we retained the additional parameter-rich ``vanilla" components, node attention and edge attention. This adjustment brings the parameter count of VFN-Diff close to that of FrameDiff (18.3 million vs. 17.4 million), facilitating a fair comparison. Moreover, node attention and edge attention are well-established practices, widely adopted in previous works such as PiFold. Consequently, these modules do not constitute the core of IPA; rather, point attention does. Hence, the replacement of point attention with the VFN-Diff approach represents a rigorously comparative implementation.

\subsubsection{Implementation of VFN-IF}~\label{sec:vfn_if}
We have adopted the PiFold framework to implement VFN, consisting of three components: the input layer, network layer, and decoder. Concerning the decoder and corresponding loss functions, VFN remains consistent with PiFold. For the network layer, we have substituted PiFold's layer, PiGNN, with the VFN layer. However, to ensure a fair comparison, the Global Context Attention module from PiFold has also been retained in VFN-IF.

\begin{wrapfigure}{r}{0.7\textwidth}
  \centering
  \includegraphics[width=0.6\textwidth]{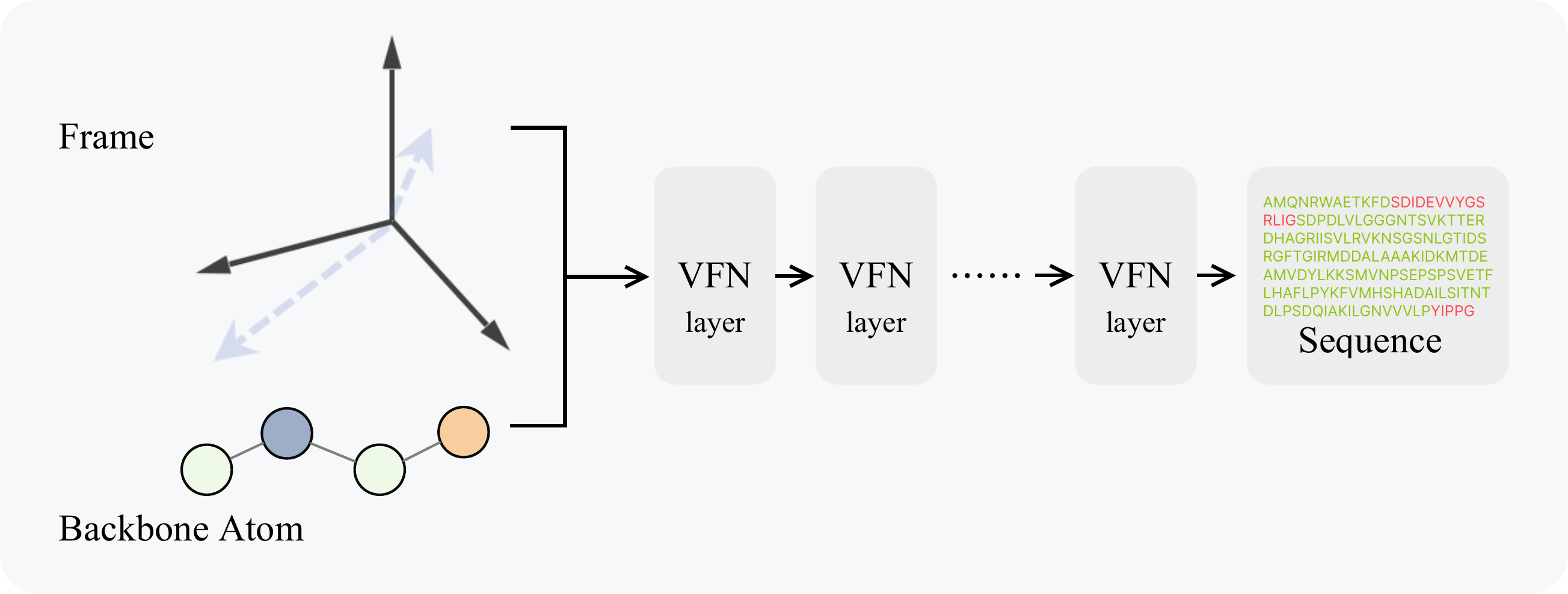}
  \caption{We harnessed the capabilities of VFN layers to effectively extract both frame and backbone atom information, facilitating the generation of sequences for inverse folding.}
  \label{fig:VFN-IF}
\end{wrapfigure}

Regarding the network input layer, PiFold employs a manual featurizer to extract interatomic features such as angles and distances. In VFN, we have completely removed this featurizer since the VFN layer itself can extract geometric features without relying on a featurizer. We initialize a portion of virtual atomic coordinates using known backbone atomic coordinates to provide the network with this information.

\subsubsection{Implementation of VFN-IFE}~\label{sec:vfn_ife}

\begin{wrapfigure}{r}{0.6\textwidth}
  \centering
  \includegraphics[width=0.5\textwidth]{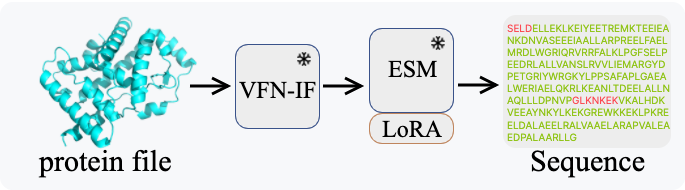}
  \caption{Flowchart depicting the fine-tuning process of frozen VFN-IF and ESM with LoRA.}
  \label{fig:VFN-IFE}
\end{wrapfigure}

In VFN-IFE, the ESM model is employed to refine the predictions made by VFN-IF. During training, the pretrained VFN-IF is frozen, and its predictions are used as input to the ESM model (as described later). The ESM model is initialized with pretrained weights and then frozen. We leverage LoRA to provide learnable weights and fine-tune the ESM. ESM is supervised with a cross-entropy loss using ground truth amino acid labels. The most critical aspect of our approach is the introduction of a relaxation method, allowing the ESM to accept probability-based inputs from VFN-IF.

For $i$-th amino acid, VFN-IF predicts a set of probabilities corresponding to 20 amino acid categories, denoted as $\vp = \{0<p_j<1\}_{j=0,...,20}$. Similarly, ESM includes word embeddings corresponding to these categories, denoted as $\rmW_{\text{ESM}} = \{\rvw_j \in \R^ {d_{\text{ESM}}}\}_{j=0,...,20}$. Therefore, we use the predicted probabilities $\vp$ from VFN to perform a weighted sum of the corresponding word embeddings $\rmW_{\text{ESM}}$ for producing the input token $s_i^{\text{ESM}}$ to ESM, denoted as:

\begin{equation}
\begin{aligned}
    s_i^{\text{ESM}} = \sum_j p_j \rvw_j
\end{aligned}
\end{equation}

\subsubsection{Local Frames of Residues}\label{sec:local_frame}

In VFN, a local frame is set up for each residue via a Gram–Schmidt process proposed by AlphaFold2  \citep{jumper2021highly}, refer to $\operatorname{rigidFrom3Points}$ algorithm in their paper. 

\subsubsection{Proof of SE(3) Invariance}
The output and the vector representation of VFN are SE(3) invariant, which are crucial for networks to attain higher performance. The proof is simple. In short, the local frame of residues is SE(3) equivariant, which ensures the invariance for the inputs of VFN and the outputs of vector field operator. Here, we provide the proof following:

\textbf{The SE(3) invariance of initial representation.} Due to all the initial representations are with respect to the local frame, the input of VFN, $\rvs_i, \rve_{i,j}, \vec{\rvq}_{k}$, is SE(3) invariant. 

\textbf{The SE(3) invariance of the vector field operator.} In our main paper, the transform matrix $T_{i\leftarrow j}$ is employed in our operators (refer to Eq.~\ref{eq:transform_op} in our main paper): 
\begin{equation}
    \rmK_j = \rmT_{i\leftarrow j} \circ \rmQ_j, \quad  \text{where} \ \  \rmT_{i\leftarrow j} = \rmT_i^{-1} \circ \rmT_j
\end{equation}

The transform matrix $T_{i\leftarrow j}$ is SE(3) invariant \textit{w.r.t.} the global reference frame, because the global frame cancels out in the computation of the transform matrix $\rmT_{i\leftarrow j}$:
\begin{equation}
\begin{aligned}
    (\rmT_{\text{global}} \circ \rmT_i)^{-1} \circ (\rmT_{\text{global}} \circ \rmT_j) &= \rmT_i^{-1} \circ \rmT_{\text{global}}^{-1} \circ \rmT_{\text{global}} \circ \rmT_j \\
                                                             &= \rmT_i^{-1} \circ \rmT_j
\end{aligned}
\end{equation}
where $\rmT_{\text{global}}$ denotes any global reference frame. Therefore, the outputs $\vec{\rvh}_{k}, \rvg_{i,j}$ of the vector field operator are SE(3) invariant.

\textbf{The conclusion of SE(3) invariance.} Finally, because the input of VFN and the vector field operator (Only these operators associated with the global reference frame) are SE(3) invariant, the intermediate variables and outputs of VFN satisfy SE(3) invariance. 

\subsection{De novo Protein Design Using VFN}
In this section, we explored \textit{de novo} protein design through the utilization of two distinct pipelines. The first pipeline involves the established flow of FrameDiff coupled with Protein MPNN, while the alternative approach adpots VFN-Diff in tandem with VFN-IFE. Both pipleines employed the ESMFold \citep{brain2022esmfold} to fold amino acid sequences into ptrotein structures.Remarkably, in one-shot experiments with a Noise Scale parameter set to 1.0, our VFN-based pipeline outperformed FrameDiff+Protein MPNN approach in terms of designability, as shown in section~\ref{sec:supp_metric}.

\begin{figure}[h]
\centering
    \includegraphics[width=1.0\linewidth]{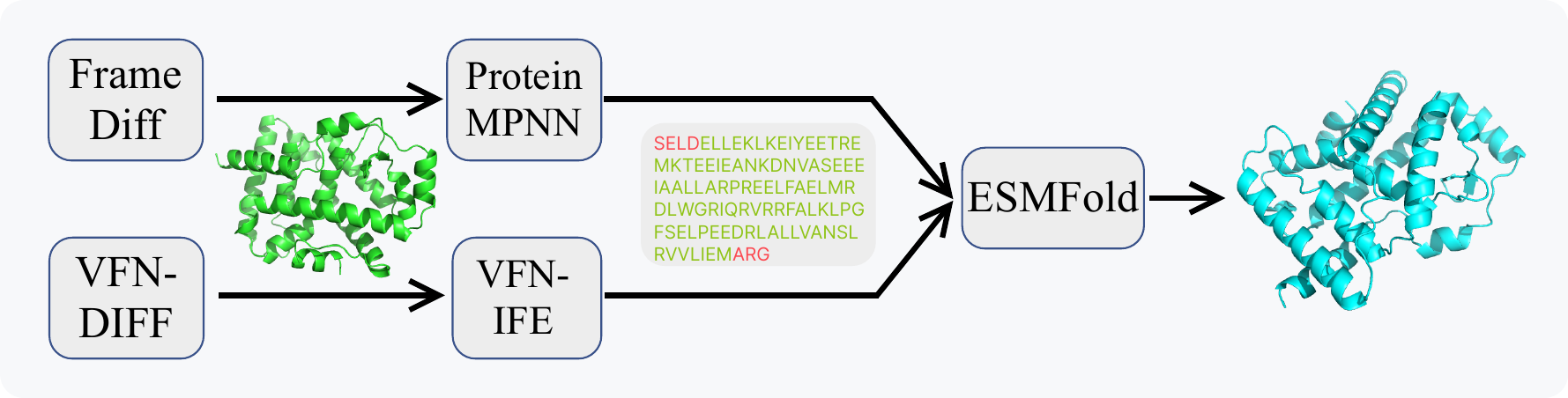}
    \caption{Whole pipelines}\label{fig:pipeline}
\label{fig:pipeline}
\end{figure}

\begin{table}[h]
\centering
\begin{tabular}{ccc}
\toprule
\textbf{} & scTM $>$ 0.5  & RMSD $<$ 2  \\
\midrule
FrameDiff + MPNN & 37.25\% & 6.88\% \\
VFN-Diff + VFN-IFE  & \textbf{45.43\%} & \textbf{11.73\%} \\
\bottomrule
\end{tabular}
\caption{The VFN-based pipeline exhibited significantly improved structural alignment compared to the FrameDiff pipeline, signifying enhanced structural consistency. Additionally, the RMSD analysis underscored the superior structural alignment achieved by the VFN-based approach.}
\label{tab:pipeline}
\end{table}

\subsection{Experiment}\label{sec:supp_experiment}
\subsubsection{Dataset Details}
\textbf{Protein diffsuion.} 
FrameDiff was re-trained using the same standards as in \citep{yim2023se}, while VFN-Diff was trained on four NVIDIA 4090 GPUs for a total duration of 10 days and 14 hours. The training dataset consisted of proteins from the PDB database        \citep{berman2000pdb} in August 2023, encompassing 21,399 proteins with lengths ranging from 60 to 512 and a resolution of $< 5$\AA.

\textbf{Inverse folding.} 
Unless specified, our experiments are conducted on the  CATH 4.2        \citep{orengo1997cath} dataset using the same data splitting as previous works such as GVP        \citep{jing2020learning} and PiFold        \citep{gao2022pifold}. The dataset consists of 18,024 proteins for training, 608 for validation, and 1120 for testing. During the evaluation, we also test our model on two smaller datasets, TS50 and TS500        \citep{jing2020learning,gao2022pifold}, to validate the generalizability. Furthermore, we also create another larger training set by incorporating data from the PDB  \citep{burley2017protein}. We apply the same strategy as in        \citep{zhou2023uni} to collect and filter structures. Additionally, the proteins with sequences highly similar to test set proteins are also removed. By using the expanded dataset, we are able to scale up the VFN-IF.

\subsubsection{Implement Details}
\textbf{VFN-Diff.} Our training regimen employs the Adam optimizer with the following hyperparameters: a learning rate of 0.0001, $\beta_1$ set to 0.9, and $\beta_2$ set to 0.999. We follow the setting of FrameDiff. For more implement details, please refer to thier paper.

\textbf{VFN-IF.} VFN-IF are trained with batch size 8 and are optimized by AdamW with a weight decay of $0.1$.  We apply a OneCycle scheduler with a learning rate of $0.001$ and train our model for a total of 100,000 iterations. 

\textbf{VFN-IFE.} 
We constructed VFN-IFE by employing a 15B ESM model in conjunction with the standard VFN-iF. The rank for LoRA applied to ESM was set to 8. All other training settings remained consistent with those of VFN-iF.

\subsubsection{Metric}\label{sec:supp_metric}  

\textbf{Protein diffusion inference.} Inferences were conducted with protein lengths ranging from 100 to 500, using a step size of 5. This resulted in the generation of 10 diffusion samples of protein at each step, with a total of 810 diffusion samples, denoted as $N_{diff}$. `Noise Scale' represents the initial noise scale in diffusion, while `Num. Step' represent the diffusion steps during inference. Protein MPNN was employed to generate `Num. Seq.' ($N_{seq}$) sequences for each sample, followed by ESMFold to create protein structure files, referred to as $N_{esm}=N_{diff} \times N_{seq}$. 

\textbf{Protein diffusion Metrics.} Metrics of protein diffusion inference experiments can be categorized into the following sections:

\begin{itemize}
    \item \textbf{Structural Similarity Metrics:} This metric evaluates the proportion of samples with a structural consensus TM score (scTM) and RMSD meeting the criteria of scTM $>$ 0.5 and scRMSD $<$ 2 Å, as indicated in table~\ref{tab:diversity_disgnability} by $\text{scTM}_{0.5}$ and $\text{scRMSD}{_2}$. In this context, scTM measures the structural similarity between ESMFold generated proteins and diffusion generated structures, while scRMSD quantifies the root-mean-square deviation in atomic positions between these structures.

    \item \textbf{Diversity:} To gauge the diversity of the generated protein sequences, we adhered to the methodology laid out in \citep{yim2023se} and leveraged MaxCluster \citep{herbert2008maxcluster} for hierarchical clustering of protein backbones. However, we opted for a higher threshold of 0.6 TM-score, in contrast to the 0.5 TM-score referenced in the literature, to impose a more rigorous clustering criterion. Furthermore, we enforced a selection criterion for cluster inclusion, requiring scTM $>$ 0.5, with the intention of mitigating the impact of proteins with low designability on diversity assessments. For each diffusion sample, a singular protein generated through ESMFold was chosen based on the highest scTM score. The diversity metric was computed as the ratio of the number of clusters to $N_{diff}$. These experimental modifications were introduced to yield results that are not only more scientifically sound but also align more closely with anticipated trends within the inference data.

    \item \textbf{pdbTM:} To assess protein novelty, we compared the ESMFold-generated PDB files, each containing the protein with the highest scTM score from a diffusion sample, to the PDB database using the Foldseek tool \citep{van2023foldseek}. The highest TM-scores of the generated samples were compared with any chain in the PDB database, and the resulting value was denoted as pdbTM. To exclude proteins with limited designability, we applied a cutoff criterion of RMSD $<$ 2, consistent with the approach used in \citep{yim2023se}. pdbTM served as a robust metric for quantifying protein novelty, reflecting the structural similarity between the generated proteins and those documented in the PDB database. In contrast to the threshold of pdbTM $<$ 0.6 employed in \citep{yim2023se}, we considered proteins with pdbTM values less than 0.7, which refers as $\text{pdbTM}_{0.7}$ in table~\ref{tab:diversity_disgnability}, as novel designs due to their substantial dissimilarity from existing proteins, resulting in the inclusion of a greater number of novel proteins in both selection.
    
\end{itemize}


\textbf{Inverse folding.} We conducted a comparative analysis involving VFN-IF, VFN-IFE, Pifold, and LM-Design. We utilized the CATH 4.2 dataset for validation and compared the results by performing ESMFold on the generated sequences (with $N_{seq} = 1$) and comparing them with the original PDB files to calculate scTM and scRMSD. This analysis was performed as a one-shot comparison.

\textbf{Whole Pipeline.} This set of experiments compared VFN-Diff + VFN-IFE with FrameDiff + ProteinMPNN. To ensure alignment of comparison standards, we used $N_{seq} = 1$ and Noise Scale = 1.0 for one-shot comparisons.

\begin{figure}[h]
    \centering
    \includegraphics[width=1\linewidth]{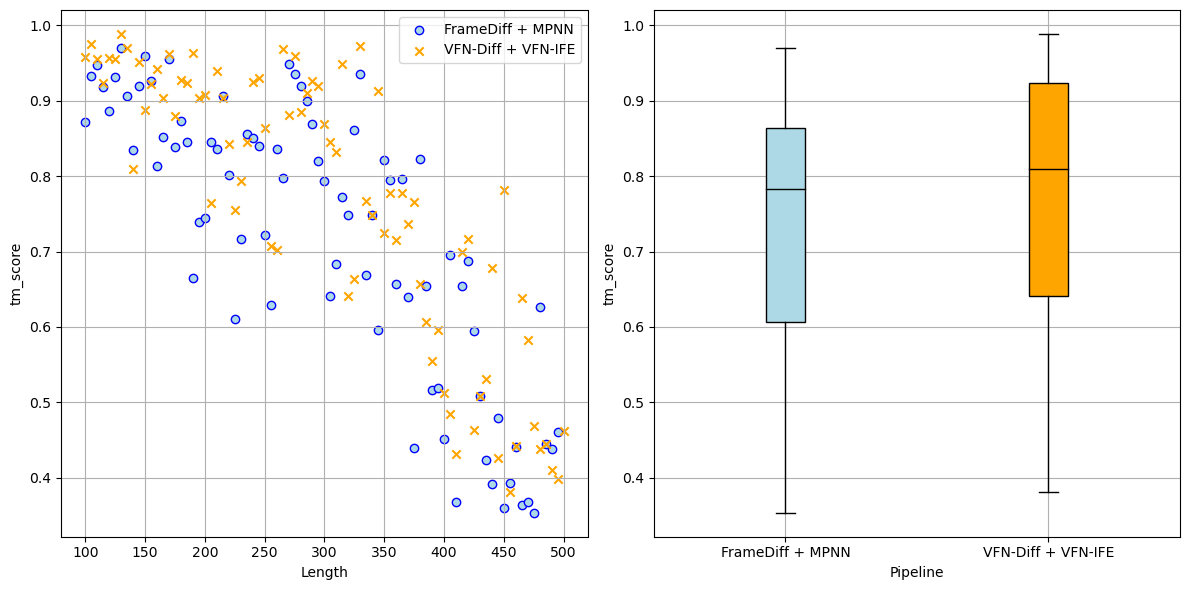}
    \caption{Whole pipeline in one-shot: FrameDiff+ProteinMPNN and VFN-Diff+VFN-IFE}
    \label{fig:enter-label}
\end{figure}

\begin{figure}[h]
    \centering
    \includegraphics[width=0.6\linewidth]{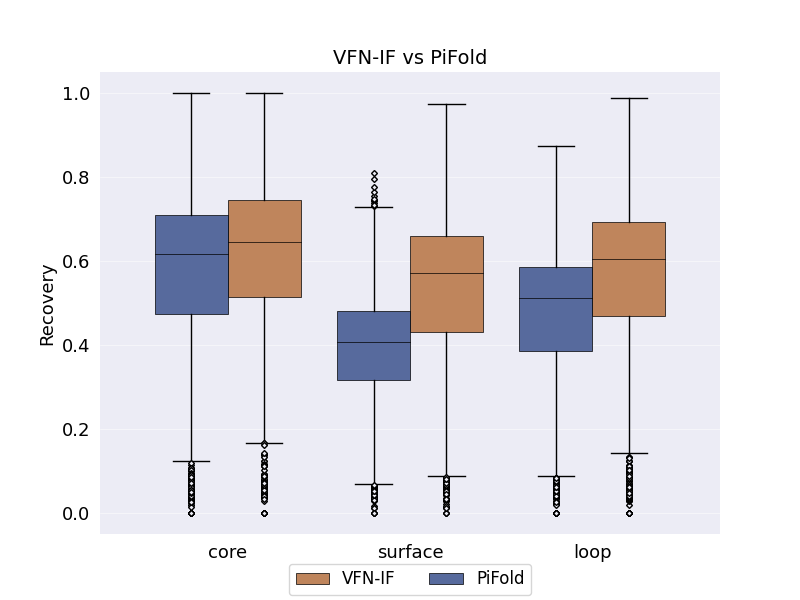}
     \caption{VFN vs PiFold on different structural contexts}
    \label{fig:core_surface_loop}
\end{figure}

\begin{table*}[h!]
    \caption{Ablation studies on the CATH 4.2 dataset. We use the default  model settings unless otherwise specified. When calculating the number of parameters, we only count the number of parameters occupied by this module in one layer. }\label{tab:ablation}
	\begin{center}
 \small 
		\subfloat[Varying the number of layers.
  \label{tab:layers}]{
              {%
 \begin{tabular}{@{}lllllll@{}}
\toprule
\multicolumn{2}{c}{\#layers}                   & 5     & 8     & 10    & 12            & 15             \\ \midrule
\multirow{2}{*}{w edge feature}   & Recovery($\%$)   & 52.53 & 54.12 & 54         & \textbf{54.3} & 54.08          \\
                                  & Perplexity & 4.3114  & 4.1986  & 4.1883 & \textbf{4.1536}           & 4.2185                 \\ \midrule
\multirow{2}{*}{w/o edge feature} & Recovery ($\%$)  & 52.68 & 53.78 & 54.28         & 54.11          & \textbf{54.7} \\
                                  & Perplexity & 4.3192  & 4.1983  & 4.1766          & 4.1829  &  \textbf{4.1687}            \\ \bottomrule
\end{tabular}
            }
		}
  \\
		\subfloat[Varying the number of vectors.  \label{tab:vec}]
		{
             {%
                \begin{tabular}{@{}llll@{}}
                \toprule
                \#Vec.      & 16    & 32             & 64    \\ \midrule
                Recovery ($\%$)  & 53.97 & \textbf{54.28} & 53.63 \\
                Perplexity & \textbf{4.1598}  & 4.1766  & 4.2636 \\ \bottomrule
                \end{tabular}
		}
            }

              
            \subfloat[Varying the $\operatorname{V-MLP}$.  \label{tab:v-mlp}]
            		{
                         {%
                           \begin{tabular}{@{}lcc@{}}
                            \toprule
                                                   & Recovery  ($\%$)     & Params        \\ \midrule
                            $\operatorname{V-MLP}$ & \textbf{54.28} & 4.2K \\
                            $\operatorname{MLP}$   &   53.42             & 36.0K         \\ 
                            w/o $\operatorname{V-MLP}$ & 53.67         & \textbf{0.0K   } \\ 
                            \bottomrule
                            \end{tabular}
            		}
                        }
            \subfloat[Varying the vector field.  \label{tab:vec_field}]
            		{
                        {%
                           \begin{tabular}{@{}lcc@{}}
                            \toprule
                                            & Recovery  ($\%$)     & Perplexity 
                                            \\ \midrule
                            Baseline       &\textbf{ 54.28} &    \textbf{4.1766}                   \\
                            w/o Vec.\  field & 35.43   &7.5455                   \\ 
                            w/o $g_{i,j}$ in Eq.~\ref{eq:edge_inter} & 53.05    &4.2907                    \\ \bottomrule
                            \end{tabular}
            		}
                        } \\
            \subfloat[Varying the transformation $\mathcal{T}$.  \label{tab:vec_field2}]
            		{
                        {%
\begin{tabular}{@{}c|ccc|cc@{}}
\toprule
      ID  & $\operatorname{Gbf}$ & $\vec{\rvh}_{k}/\Vert \vec{\rvh}_{k} \Vert$ & $\operatorname{Transformation}$   & Recovery(\%) &Perplexity \\ \midrule
1     & \ding{51} & \ding{51} & \ding{51} & \textbf{ 54.28} &\textbf{4.1766}\\
2       & \ding{55} & \ding{51} & \ding{51} & 53.55 &4.2495\\
3 & \ding{51} & \ding{55} & \ding{51} & 53.36 &4.2238\\
4   & \ding{55} & \ding{55} & \ding{55} & 52.89 &4.3268 \\ \bottomrule
\end{tabular}
            		}
                        }
	\end{center}
\end{table*}

\subsubsection{Ablation study} 

\subsection{Ablations}  
In this section, we carefully investigate the design choices of vector field modules proposed here based on CATH 4.2 (inverse folding). 

\noindent\textbf{The number of layers.} 
We investigate the impact of modifying the number of layers on recovery and perplexity in Table \ref{tab:layers}. Increasing the number of layers from 5 to 15 results in a marginal improvement in recovery scores, with the highest recovery achieved at 54.72\% for 12 layers. Ablation experiments without the edge featurizer show that with or without edge features, the performance is comparable. Especially when the number of layers reaches 15, it even achieves better results, indirectly proving the effectiveness of VFN-IF and its potential for reducing reliance on hand-crafted features.

\noindent\textbf{The number of vectors.} 
In Table \ref{tab:vec} we also perform an ablation study to  determine the optimal number for vectors. It demonstrates that increasing the vector number from 16 to 32  brings a slight improvement in recovery score from 53.72\% to 54.26\%, while maintaining a low perplexity score of 4.14. However, further increasing the number to 64 results in a decrease in recovery score to 53.83\% and a higher perplexity score of $4.20$.  Overall, these findings suggest that 32 is a more suitable choice.

\noindent\textbf{V-MLP.} In Table \ref{tab:v-mlp}, we observe that using $\operatorname{V-MLP}$ outperforms using a regular $\operatorname{MLP}$ or not using it at all. Compared to using a regular $\operatorname{MLP}$, $\operatorname{V-MLP}$ significantly reduces the parameter count.

\noindent\textbf{Vector field design.}
Results in Table~\ref{tab:vec_field} show that removing the whole vector field operator leads to a significant drop in recovery, indicating its importance in capturing protein folding patterns. The incorporation of the $g_{i,j}$ in the edge aggregation module also has a substantial effect on the performance of the model.

\noindent\textbf{Vector field operator design.} We validate the effectiveness of $\operatorname{RBF}$ and $\vec{\rvh}_{k}/\Vert \vec{\rvh}_{k} \Vert$ in Eq.~\ref{eq:h_to_g}, as shown in Table   \ref{tab:vec_field2}. Furthermore, it illustrates that completely excluding the Eq.~\ref{eq:h_to_g} and directly flattening the vectors $\rmH_{i,j}$ into $\rvg_{i,j}$ leads to an obvious performance decrease.

\subsubsection{Visualization results} 
\begin{figure}[h]
\centering
    \includegraphics[width=1.0\linewidth]{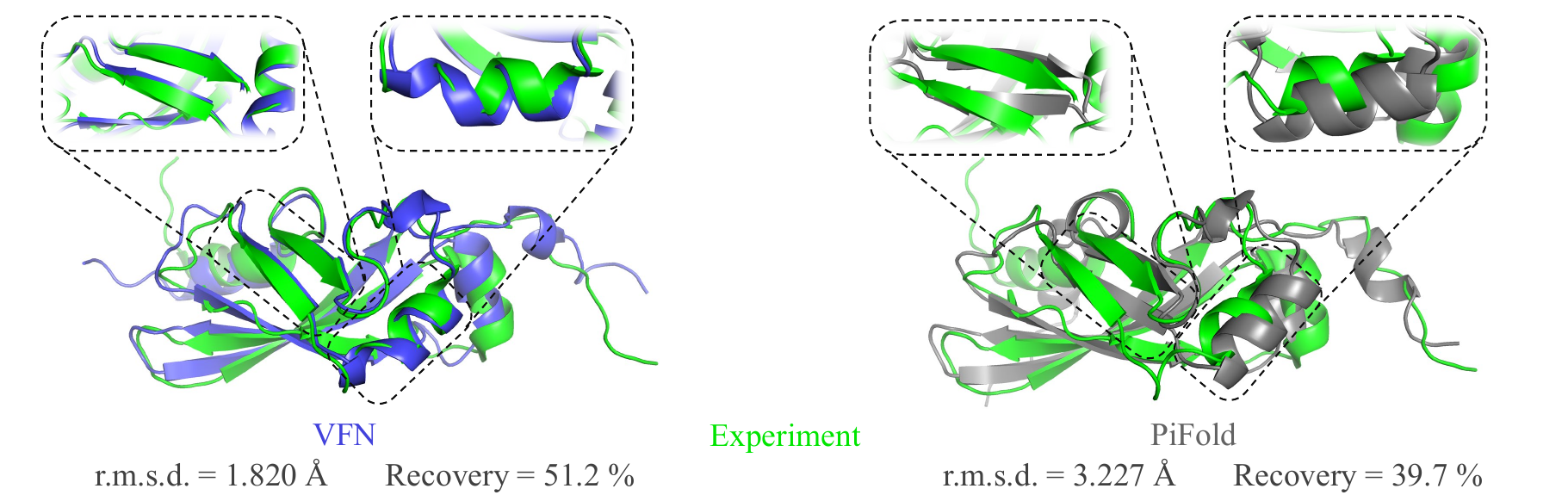}
   \caption{
   Visualization results of a challenging sample (PDB 2KRT).  We use AlphaFold2 to recover the structure based on the predicted sequence and compare it 
   against 
   the experimentally determined ground-truth  structure. 
   }
\label{fig:2krt}
\vspace{-1.0em}
\end{figure}

\begin{figure}[h]
\centering
    \includegraphics[width=1.0\linewidth]{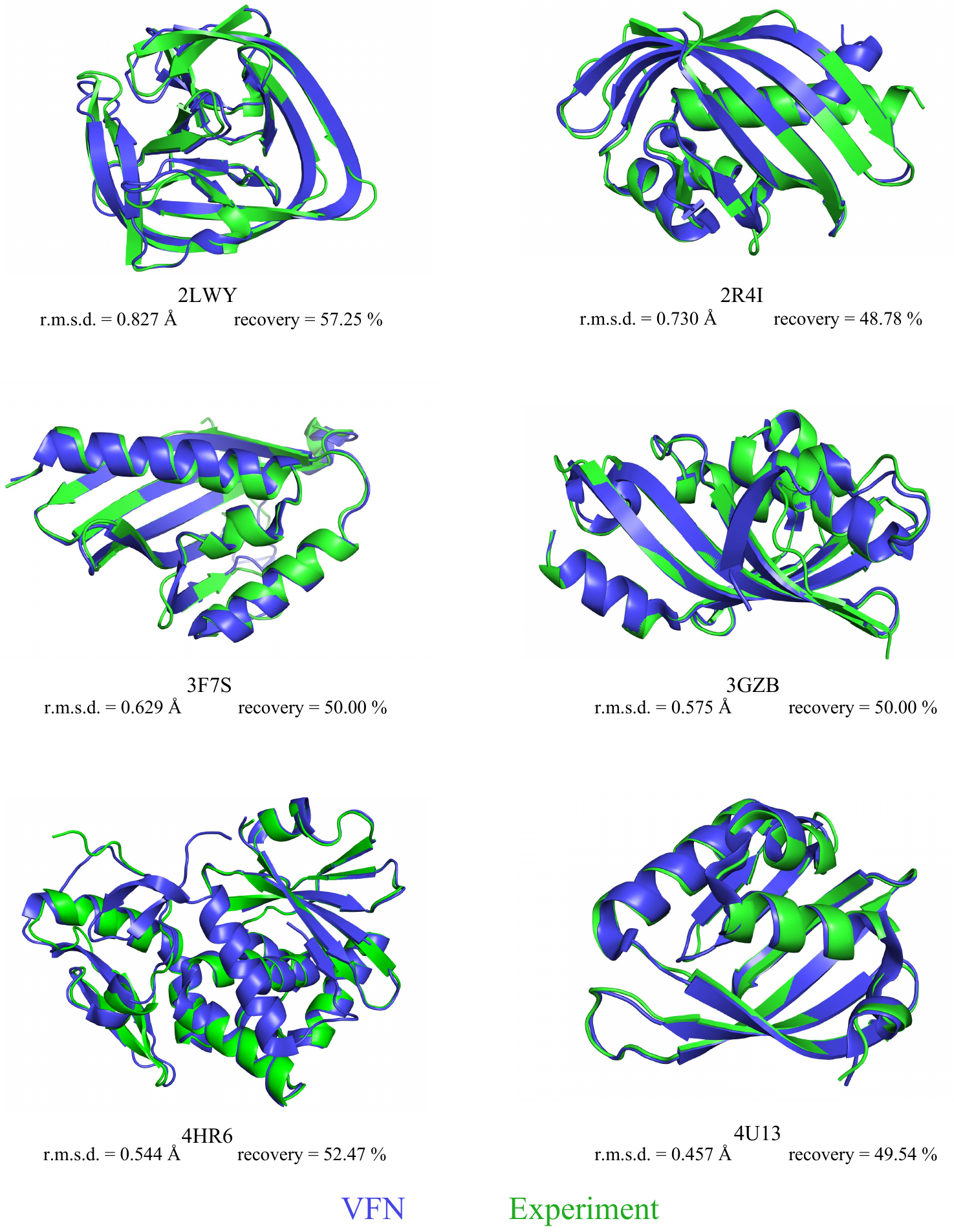}
    \caption{The visualization result of VFN.}\label{fig:vis2}
\label{fig:vis3}
\end{figure}

In this section, we visualize some of the protein structure restoration results of VFN-IF, as shown in figure~\ref{fig:vis1}, figure~\ref{fig:vis3} and figure~\ref{fig:2krt}.

\begin{figure}[h]
\centering
    \includegraphics[width=1.0\linewidth]{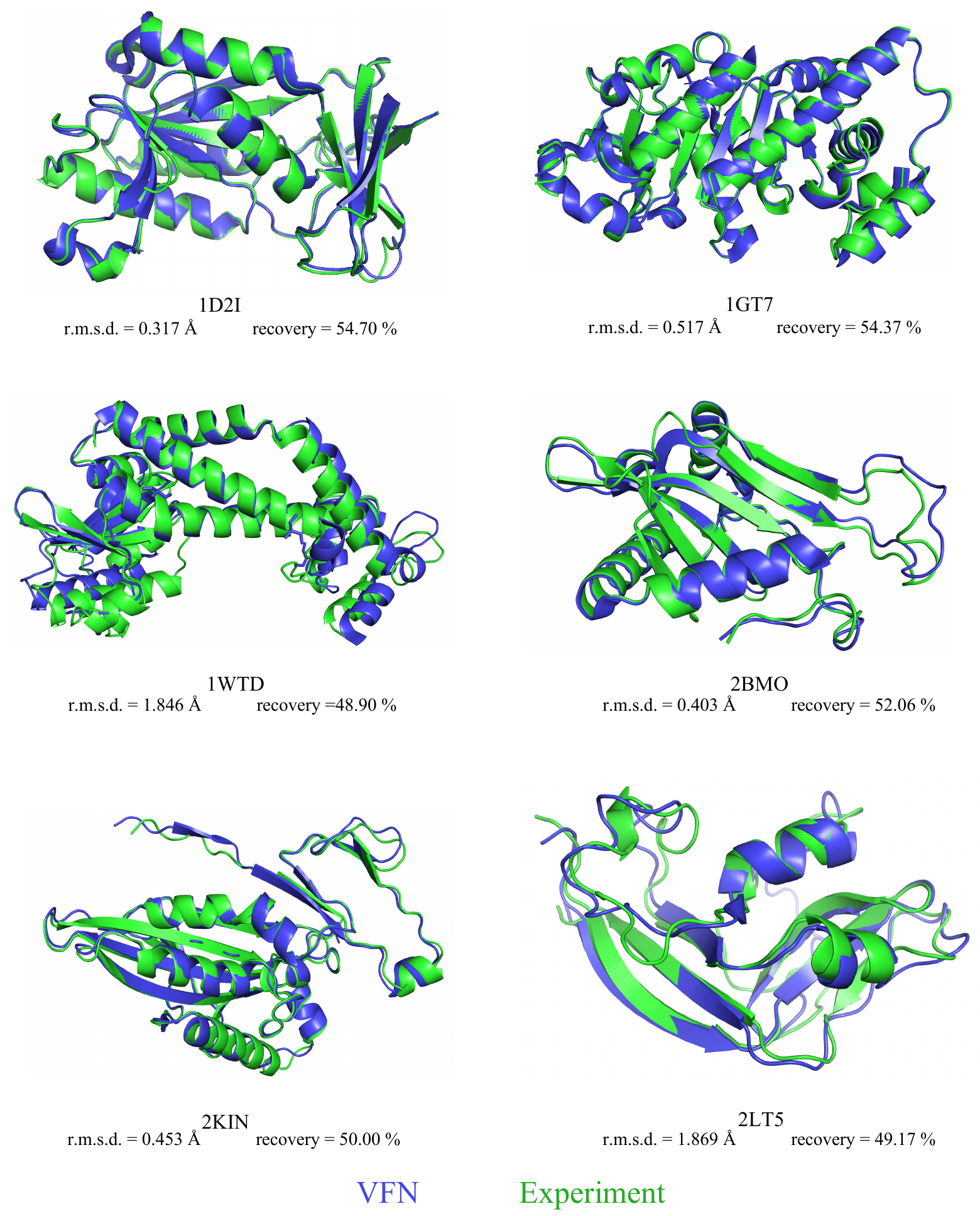}
    \caption{The visualization result of VFN.}\label{fig:vis1}
\label{fig:vis1}
\end{figure}

\end{document}